\documentclass[aps,prd,showpacs,preprintnumbers,twelvepoint,floatfix]{revtex4}
\usepackage{epsfig}
\usepackage{latexsym,amsmath,amssymb,amstext}
\usepackage{float}
\usepackage{graphicx}
\usepackage{color}

\newcommand\bef{\begin{figure}}
\newcommand\eef[1]{\label{fg:#1}\end{figure}}
\newcommand\beq{\begin{equation}}
\newcommand\eeq[1]{\label{#1}\end{equation}}
\newcommand\beqa{\begin{eqnarray}}
\newcommand\eeqa[1]{\label{#1}\end{eqnarray}}
\newcommand\bet{\begin{table}}
\newcommand\eet[1]{\label{tb:#1}\end{table}}

\newcommand\fgn[1]{Figure \ref{fg:#1}}

\newcommand\scn[1]{Section \ref{sec:#1}}
\newcommand\apx[1]{Appendix \ref{sec:#1}}
\newcommand\tbn[1]{Table \ref{tb:#1}}

\newcommand\ie{{\sl i.e.\/}}

\newcommand\etal{{\sl et al.\/}}
\newcommand\jhep{{\sl J.\ H.\ E.\ P.\/}\ }
\newcommand\np{{\sl Nucl.\ Phys.\/}\ }
\newcommand\npps{{\sl Nucl.\ Phys.\ Proc.\ Suppl.\/}\ }

\newcommand\plt{{\sl Phys.\ Lett.\/}\ }

\newcommand{\alphas}{\alpha_{\scriptscriptstyle S}} 
 
\newcommand{\dsps}{\Delta_S}
\newcommand{\dvav}{\Delta_V}
\newcommand{\nmd}{{N_{\scriptscriptstyle MD}}}
\newcommand{\prp}[1]{C^{(+#1)}}
\newcommand{\prm}[1]{C^{(-#1)}}
\newcommand{\WLS}{\chi_{\scriptscriptstyle L}}

\newcommand{\mn}{{\rm min}}
\newcommand{\mx}{{\rm max}}
\newcommand{\IR}{{\scriptscriptstyle IR}}
\newcommand{\M}{{\scriptscriptstyle M}}
\newcommand{\MSbar}{{\overline{\scriptscriptstyle MS}}}
\newcommand{\N}{{\scriptscriptstyle N}}
\newcommand{\PS}{{\scriptscriptstyle PS}}

\newcommand{\UV}{{\scriptscriptstyle UV}}

\begin{document}
\title{Hadronic Screening in Improved Taste}
\author{Sourendu\ \surname{Gupta}}
\email{sgupta@theory.tifr.res.in}
\affiliation{Department of Theoretical Physics, Tata Institute of Fundamental
         Research,\\ Homi Bhabha Road, Mumbai 400005, India.}
\author{Nikhil\ \surname{Karthik}}
\email{nikhil@theory.tifr.res.in}
\affiliation{Department of Theoretical Physics, Tata Institute of Fundamental
         Research,\\ Homi Bhabha Road, Mumbai 400005, India.}

\begin{abstract}
We present our results on meson and nucleon screening masses in finite
temperature two flavour QCD using smeared staggered valence quarks and
staggered thin-link sea quarks with different lattice spacings and quark
masses. We investigate optimization of smearing by observing its effects
on the infrared (IR) and ultraviolet (UV) components of gluon and quark
fields. The application of smearing to screening at finite temperature
also provides a transparent window into the mechanism of the interplay
of smearing and chiral symmetry.  The improved hadronic operators show
that above the finite temperature cross over, $T_c$, screening masses are
consistent with weak-coupling predictions.  There is also evidence for
a rapid opening up of a spectral gap of the Dirac operator immediately
above $T_c$.
\end{abstract}

\pacs{12.38.Mh, 11.15.Ha, 12.38.Gc}
\preprint{TIFR/TH/13-01}
\maketitle

\section{Introduction}\label{sec:intro}

Screening masses control finite volume effects at finite temperature in
equilibrium. Studies of the final state of fireballs produced in heavy-ion
collisions indicate that they are near equilibrium. So the study of
screening masses a little below the QCD cross over temperature, near the
freeze out, should improve our understanding of experimental conditions.
In addition, the vector screening masses below $T_c$ should be of direct
relevance to the study of mass spectra of dileptons and photons.

There are also interesting questions about the nature of the high
temperature phase which are addressed by a study of screening masses. In
QCD at temperatures, $T$, of a few times the crossover temperature, $T_c$,
analysis of the weak coupling series in powers of the gauge coupling, $g$,
indicates that the physics of the magnetic scale of momentum, $g^2T$, is
potentially non-perturbative.  As a result, it may be possible to find
phenomena in hot QCD, only involving harder scales, which are amenable
to a suitable weak coupling analysis.  For example, the fermionic part
of the pressure, as well as its derivatives with respect to chemical
potentials, the quark number susceptibilities (QNS), seem to admit
reasonably accurate weak-coupling descriptions at temperatures of $2T_c$
or above \cite{qcdft}.

However, even among static fermionic quantities, screening masses
(the inverses of screening lengths) present a confused picture. Most
computations have been performed with staggered quarks, and these seem to
indicate that there are strong deviations from weak coupling prediction
\cite{kogut,fftmass,quasi,cheng,edwin}. On the other hand, computations
with Wilson quarks give results which are closer to free field theory
\cite{brandt}, although they deviate in detail from predictions of weak
coupling theory \cite{dr,htl}.  Since the same pattern is visible in
the quenched theory \cite{scrnp}, we can attribute the major part of the
discrepancy to valence quark artifacts.  Here we examine this question
systematically using staggered sea quarks and improved staggered valence
quarks. Indeed, we see that smeared valence quarks provide a significant
improvement.  Using these we find that a weak coupling expansion does
work quantitatively for the description of fermionic screening masses
at finite temperature. In addition, our results may constrain models of
thermal effects on hadrons below and close to the QCD cross over.

A significant technical component of this work is the exploration
of the cause of improvement in lattice measurements when smeared
gauge fields are introduced into the staggered quark propagators
\cite{APE,HYP,ST,HEX}. Smeared operators have been explored extensively
in the literature earlier \cite{taste}. Here we explore optimization
of smearing parameters by direct observation of the effects on UV
and IR modes separately. It also turns out that the application to
finite temperature provides a transparent window into the interplay of
improvement and chiral symmetry.

Discussion of technical lattice issues in this paper are confined to
the next two sections.  Readers who are interested only in the results
for thermal physics can read the last two sections.

\section{Methods and Definitions}\label{sec:method}

We generated configurations for the Wilson gauge action and two flavours
of thin-link staggered sea quarks using the R-algorithm.  For $am=0.015$
we used lattice sizes $N_t\times N_s^3$ with $N_t=4$ and $N_s=8$, 12, 16
and 24 for finite $T$ studies, and scanned a range of gauge couplings,
$\beta$, to find the cross-over coupling $\beta_c$. This is completely
standard, and the results are collected in \apx{tc}.  The simulations were
done using a MD time step $dt=0.01$ and trajectories with number of steps,
$\nmd=100(N_s/8)$. We checked that halving the time step did not change
the results. We observed that it was sufficient to discard the first three
hundred trajectories for thermalization. The configurations analyzed were
thermally equilibrated and spaced one autocorrelation time apart. Details
of the runs and statistics are collected in \tbn{confnew}. This data
set is called the set N in the rest of this paper.

We also studied configurations generated earlier along a line of constant
$m_\pi$ with $N_t=4$ defined by setting $am=0.025$ at the corresponding
$\beta_c=5.2875$ \cite{nt4} and $N_t=6$ defined by the choice $am=0.01667$
at its $\beta_c=5.425$ \cite{nt6}. The data from \cite{nt4} is referred
to as set O in this paper, and the data of \cite{nt6} as set P. Hadronic
screening masses from the data set P have been reported earlier using
thin-link staggered valence quarks \cite{quasi}; its inclusion in this
study enables a clear understanding of the effects of smearing. 

\bet
\begin{center}
\begin{tabular}{|l||r@{.}l|c|l||r@{.}l|r|r||r@{.}l|r|r|}
\hline
$\beta$
   & \multicolumn{4}{|c||}{$T=0$, $16^4$}
   & \multicolumn{4}{|c||}{$4\times16^3$}
   & \multicolumn{4}{|c|}{ $4\times24^3$}\\ 
\cline{2-13}
   & \multicolumn{2}{c|}{$am$} & $P$ & \multicolumn{1}{c||}{$T/T_c$}
   & \multicolumn{2}{c|}{$am$} & \multicolumn{1}{c|}{$\tau$} & $N$ 
   & \multicolumn{2}{c|}{$am$} & \multicolumn{1}{c|}{$\tau$} & $N$ \\
\hline
5.25  & 0&0165&0.4790 (3) & 0.92 (1) & 0&0165&19 &{65}&\multicolumn{2}{c|}{} & &  \\
5.26  & 0&0160&0.4827 (4) & 0.96 (1) & 0&0160&31 &{51}&\multicolumn{2}{c|}{} & &  \\
5.27  & 0&0153&0.4860 (5) & 0.98 (1) & 0&015 &72 &{48}&\multicolumn{2}{c|}{} & &  \\
5.2746& 0&015 &0.4873 (4) & 1.00     &\multicolumn{2}{c|}{}& &       &\multicolumn{2}{c|}{} & &  \\
5.275 & 0&015 &0.4873 (5) & 1.01 (1) & 0&015 &328&{76}&\multicolumn{2}{c|}{} & &  \\
5.28  & 0&0146&0.4887 (6) & 1.02 (1) & 0&015 &65 &{62}&\multicolumn{2}{c|}{} & &  \\
5.29  &\multicolumn{2}{c|}{}&& 1.06 (1) & 0&015 &21 &{49}&\multicolumn{2}{c|}{} & &  \\
5.3   & 0&0138&0.4957 (7) & 1.10 (1) & 0&0138& 8 &{59}&\multicolumn{2}{c|}{} & &  \\
5.335 &\multicolumn{2}{c|}{}&& 1.20 (1) & 0&0125& 7 &{75}&\multicolumn{2}{c|}{} & &  \\
5.34   & 0&0115 & 0.5100 (2) & 1.29 (3) &\multicolumn{2}{c|}{} & & &0&0115&$\ \ 6$&{50}\\
5.38  & 0&01  &0.5243 (1) & 1.51 (5) &\multicolumn{2}{c|}{}  & & &0&01  &$\ 6$&{57}\\
5.48  & 0&0075&0.5480 (2) & 2.03 (9) &\multicolumn{2}{c|}{}  & & &0&0075&3&{79}\\
\hline
\end{tabular}
\end{center}
\caption{The number of independent configurations, $N$, obtained with the
 coupling, $\beta$, the bare quark mass, $am$, and the autocorrelation time,
 $\tau$, for that simulation. Also given are the plaquette value, $P$,
 measured at $T=0$, and the temperature, $T/T_c$, inferred from it.}
\eet{confnew}

We studied screening correlators of mesons and the nucleon.  The valence
quarks were improved using one level smeared gauge links \cite{APE,
HYP, ST, HEX}; the optimization of the smearing algorithm is discussed
in \scn{smear}. In the course of this study we needed to estimate
the extremal eigenvalues of the staggered Dirac operator. This was
done using a Lanczos iteration \cite{golub}. The tridiagonal matrix
generated using this process was diagonalized using the Lapack routine
DSTEVX. The investigation of smearing also needed the determination
of the taste partners of the pion. For all the correlation functions
we used Coulomb gauge fixed wall-sources to project on the modes with
vanishing spatial momentum.  At $T=0$ and for temperatures below $T_c$,
multiple wall-sources separated by 4 lattice units were used. We checked
that these gave statistically independent results; an observation that
could be justified after the fact by the measurement of the Goldstone
(local) pion mass.

The screening correlator for the meson $\gamma$ was parametrized as
\beq
 C_\gamma(z)=A_\gamma\cosh\left[\mu_\gamma\left(\frac{N_s}2-z\right)\right]
      +(-1)^z A_\gamma'\cosh\left[\mu_\gamma'\left(\frac{N_s}2-z\right)\right].
\eeq{msn}
The alternating component is absent for the Goldstone pion
\cite{golterman}.  Among local operators, we measured the scalar (S)
corresponding to the $\sigma/a_0$ meson at $T=0$, the pseudoscalar (PS)
corresponding to the $\pi$ at $T=0$, the vector (V, $\rho$ at $T=0$)
and the axial vector (AV). At $T=0$ all three polarizations of the V and
AV are equivalent. However, for $T>0$, we need to distinguish between
the spatial ($V_s$, $AV_s$) and temporal ($V_t$, $AV_t$) polarizations.
For the study of taste symmetry we also measured the non-local taste
partners in some of these channels.  Following \cite{golterman}, the
nucleon correlator is parametrized as
\beqa
\nonumber
  C_\N(z) &=& A_\N\left\{\exp\left[\mu_\N\left(\frac{N_s}2-z\right)\right]
          + (-1)^z \exp\left[-\mu_\N\left(\frac{N_s}2-z\right)\right] \right\}\\
   &&\quad+A_\N'\left\{(-1)^z \exp\left[\mu_\N'\left(\frac{N_s}2-z\right)\right]
          + \exp\left[-\mu_\N'\left(\frac{N_s}2-z\right)\right]\right\}.
\eeqa{nuc}
The screening masses, $\mu_\gamma$, $\mu_\N$, and the remaining parameters
were extracted from the measured correlators by fitting to the above
forms.  The covariance between the measurements at different $z$ were
taken care of in the fits.  The mean and the error of the parameters
were estimated by bootstrap.  Fits were made to the ranges $z_\mn\le
z\le z_\mx$, where $z_\mx$ was never more than two sites from the middle
of the lattice, $z_\mn$ was never less than two sites from the source,
and the number of data points used was always greater than the number of
parameters being fitted.  Among the fits satisfying $\chi^2/\rm{DOF}<2$,
we chose as the reported estimate of the parameter and its error to
be that which was consistent with the smallest $\mu$ within 2$\sigma$
and had the smallest error.  Chiral symmetry restoration can be tested
through the mass splittings
\beq
\dsps=\mu_S-\mu_{PS}, \quad\quad \dvav=\mu_{AV_s}-\mu_{V_s},
\eeq{spsvav}
as well as the parity projected correlators $C^{\pm\gamma}=C^\gamma
\pm(-1)^zC^{\gamma'}$, where the mesons $\gamma$ and $\gamma'$ are parity
partners \cite{quasi}.

We computed quantities in a fermionic free field theory (FFT) by numerical
inversion of the fermion matrix on a trivial gauge configuration (all
links being the unit matrix). These quark propagators were then subjected
to exactly the same analysis as in the interacting theory. The negative
chiral projections of the screening correlators, $C^{-\gamma}$, vanish
in the chirally symmetric phase, and the approach to FFT can be studied
using the positive chiral projections, $C^{+\gamma}$.

\section{Study of smearing}\label{sec:smear}

\bef
\begin{center}
\includegraphics[scale=0.35]{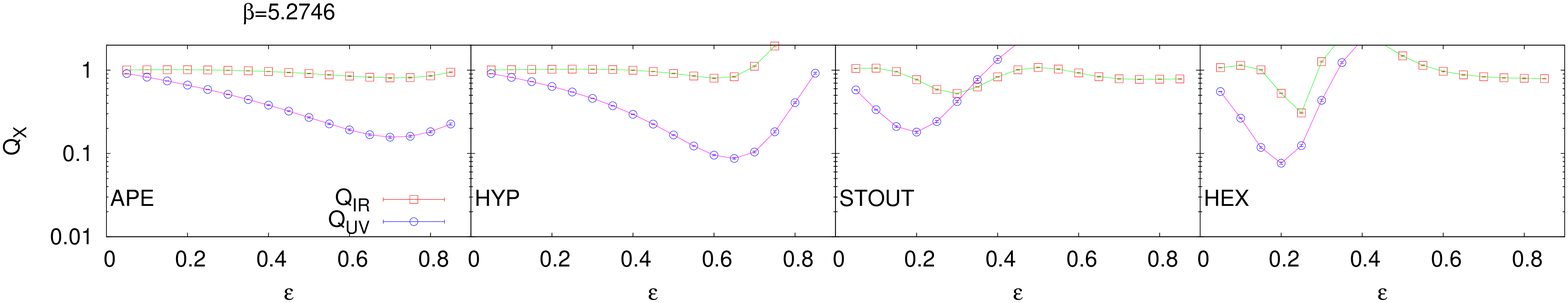}
\includegraphics[scale=0.35]{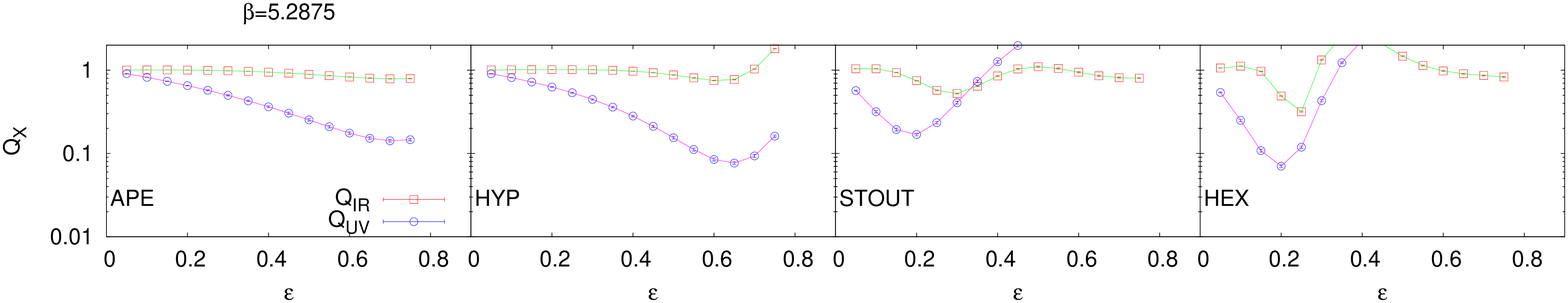}
\includegraphics[scale=0.35]{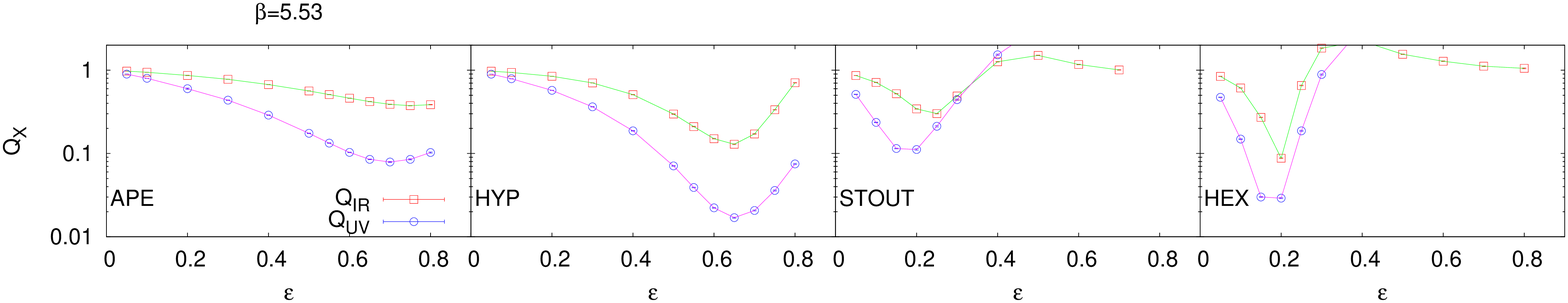}
\end{center}
\caption{Power suppression, $Q$, in the UV and IR with different kinds of
 smearing at three different lattice spacings for $T=0$. A halving of the
 lattice spacing leads to a weak change in the optimal value of $\epsilon$.}
\eef{Q}

\bef
\begin{center}
\includegraphics[scale=0.65]{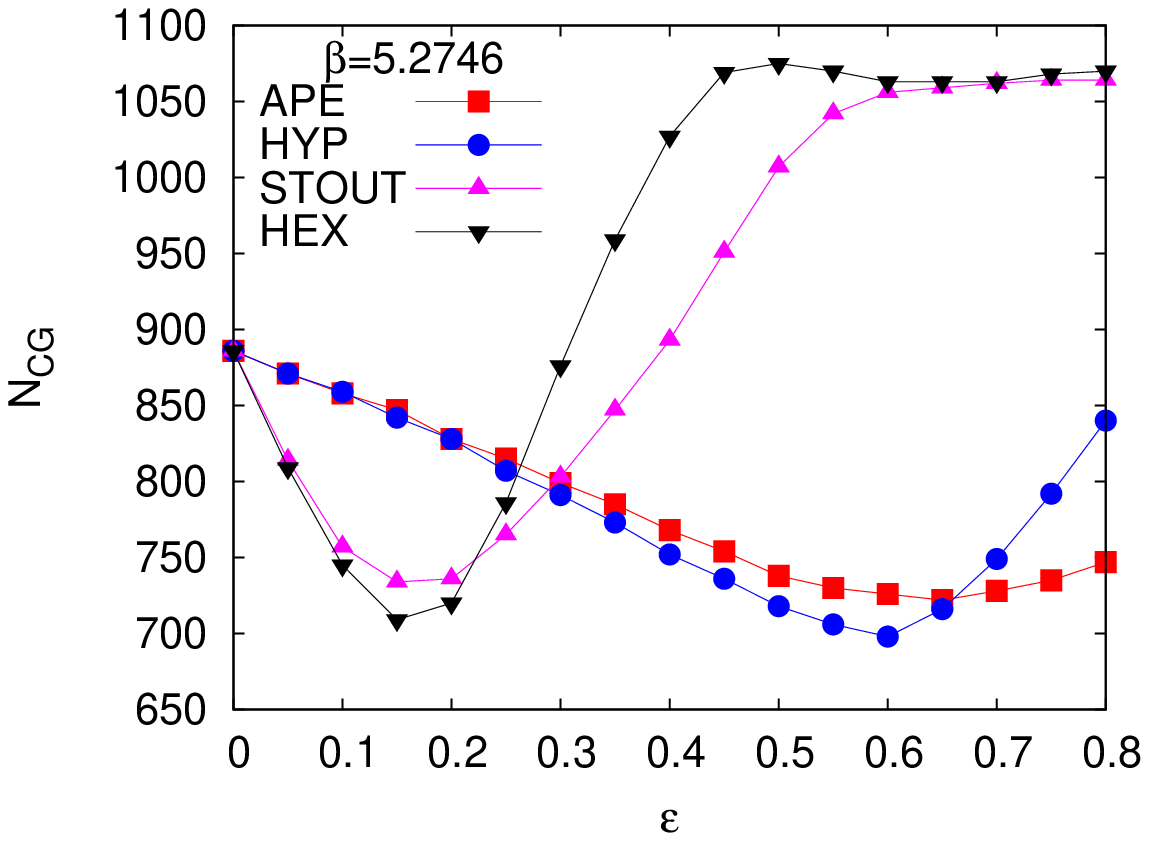}
\includegraphics[scale=0.65]{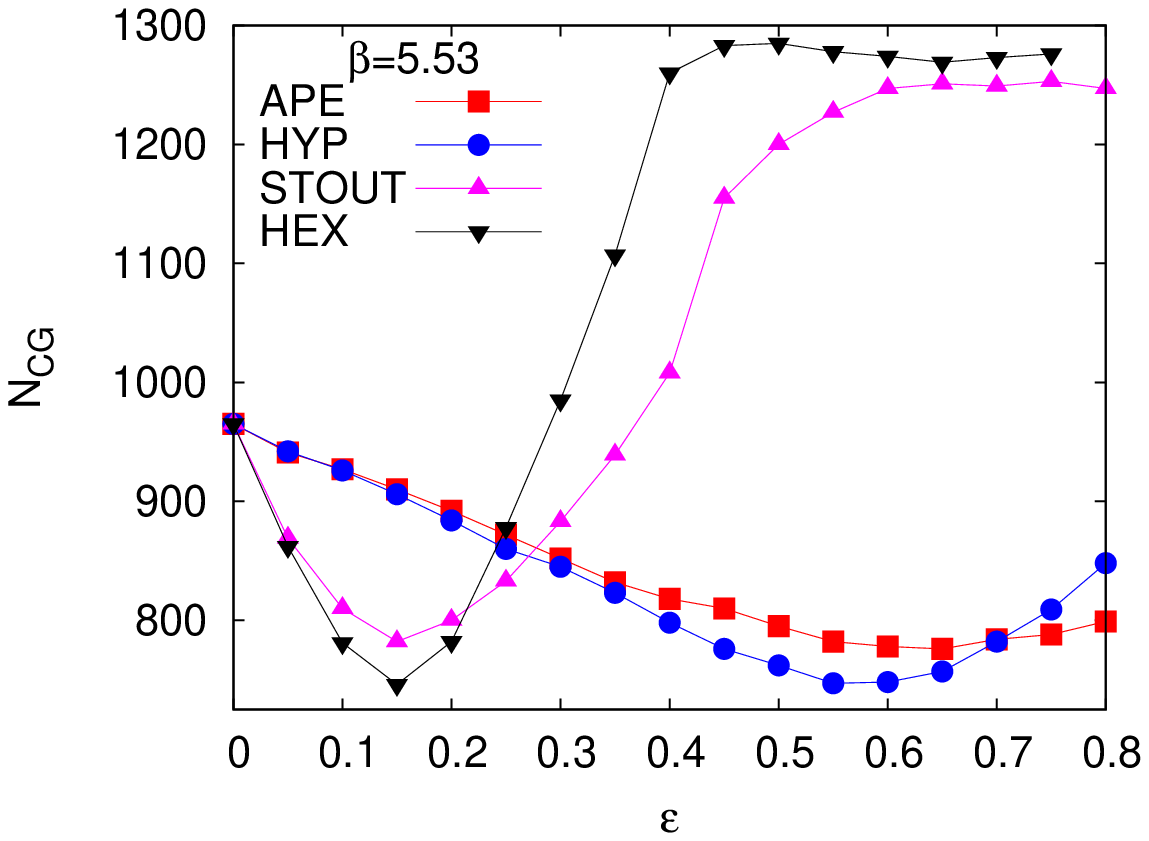}
\includegraphics[scale=0.65]{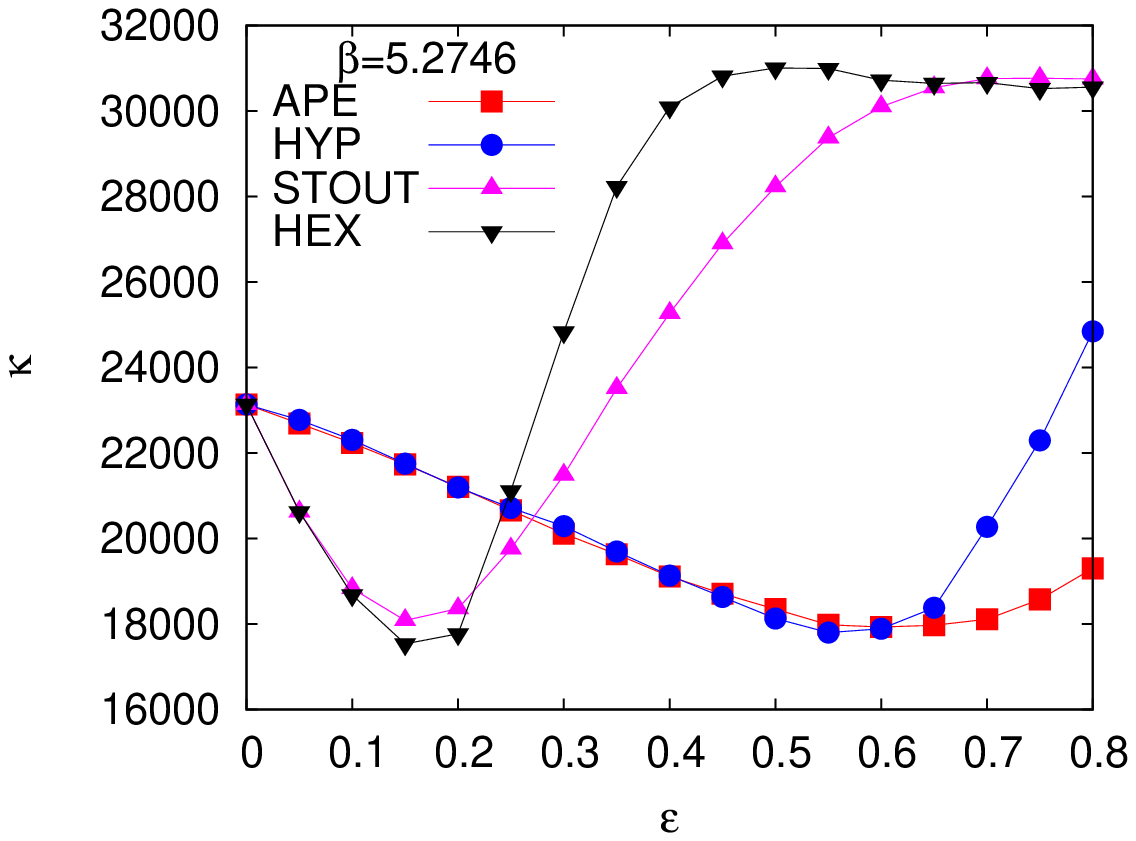}
\includegraphics[scale=0.65]{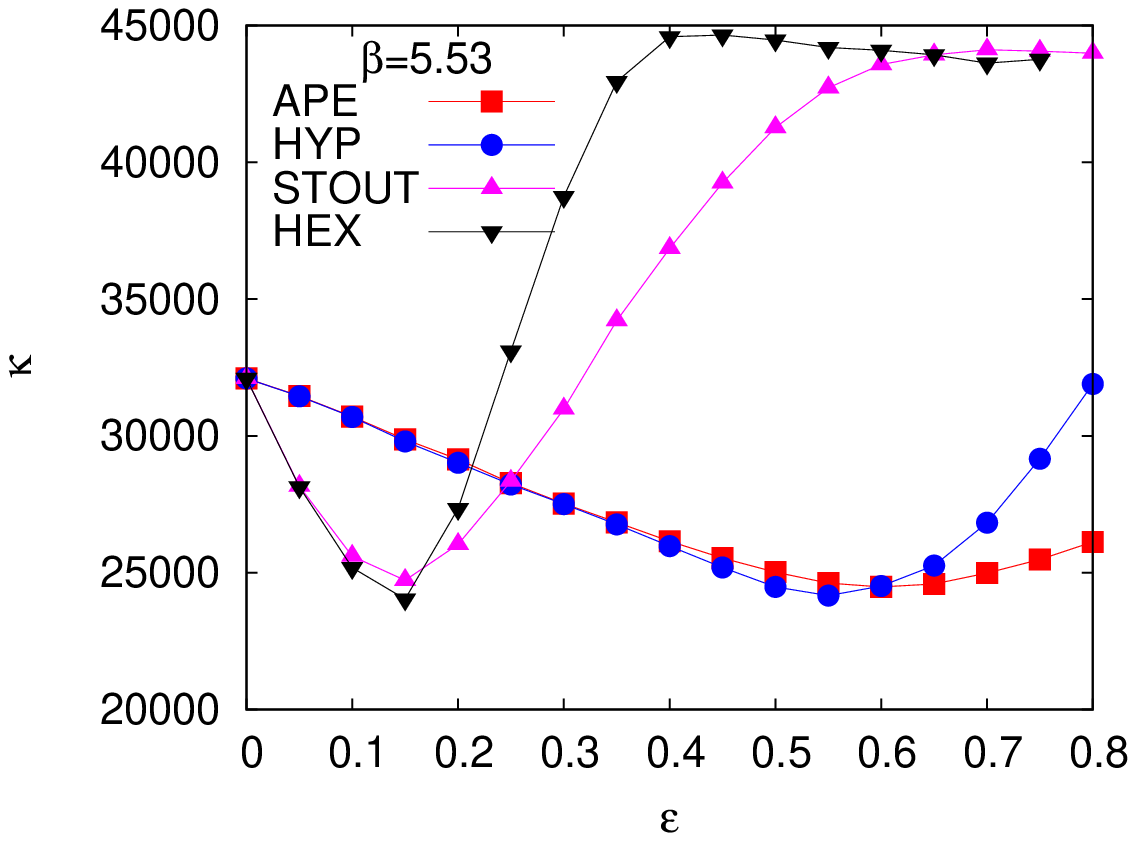}
\includegraphics[scale=0.35]{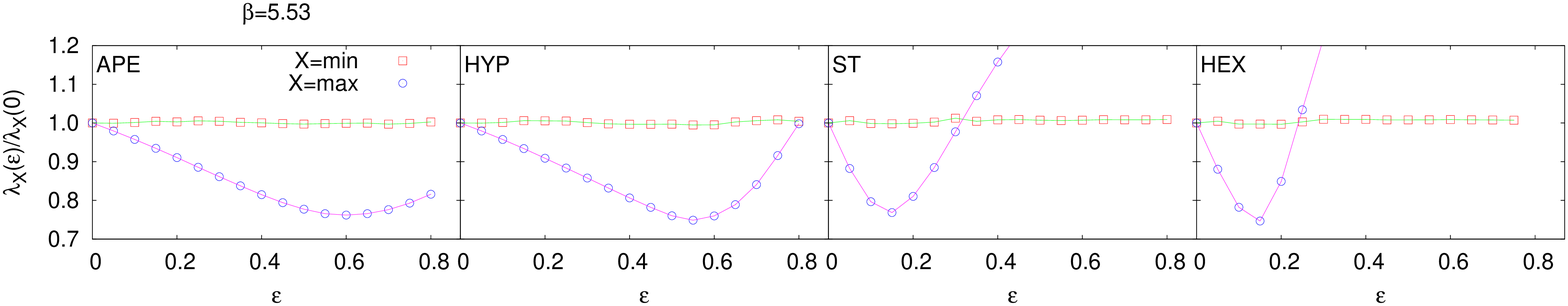}
\end{center}
\caption{Speedup of the conjugate gradient inversion for $T=0$ at two different
lattice spacings as a function of $\epsilon$ in the various different
smearing schemes.  The optimal value of $\epsilon$ in each smearing
scheme agrees with that seen in the glue sector.  This closely follows
the change in the condition number, $\kappa$, of the fermion matrix.
The last panel shows that the change in the condition number comes from
the UV, \ie, $\lambda_{max}$; the IR, \ie, $\lambda_{min}$, is almost
unchanged by smearing.}
\eef{cgkno}

Smeared gauge links improve the scaling behaviour of staggered quarks
\cite{taste}. This has been measured through staggered pion taste
splitting. Optimal parameter values have been obtained numerically and
there have been attempts to understand the results in weak-coupling
theory \cite{HEX}.

We examined four schemes which are currently popular: APE \cite{APE},
HYP \cite{HYP}, Stout \cite{ST}, and HEX \cite{HEX}.  All these schemes
involve replacing the gauge field on a link by a weighted sum of gauge
transporters over different paths connecting the end points of this
link. The more steps of such smearing we take, the more non-local the
action becomes. In order to retain a degree of locality compatible with
the sea quark action, we restricted ourselves to one step of smearing.
The APE and Stout schemes have a single free parameter, $\epsilon$,
which determines how much importance is given to link neighbours. The
HYP and HEX schemes have three different fattening parameters in three
orthogonal directions.  We restricted our study to the subset which
have equal contributions from all directions, controlled by a single
parameter $\epsilon$.

\subsection{Optimization of smearing parameters}\label{sec:optim}

The usual lore about smearing is that it suppresses the dependence
of operators on high-momentum field modes. Since the lattice cutoff
affects UV modes strongly, the result could be closer to the continuum
limit. Since field operators have a gauge dependence, it is hard to
test this idea directly on gauge fields. Instead we tested it on the
plaquette at a site $x$ averaged over all 6 orientations, $P(x)$.
As for any local operator, one can work with the Fourier transform,
$P(k)$, and the power spectrum, $E(k)$, where
\beq
 P(k)=\sum_x\exp\left(ik\cdot x\right)P(x) \qquad{\rm and}\qquad
 E(k)=|P(k)|^2,
\eeq{four}
the mode numbers $k_\mu = \pi(2\ell_\mu+\zeta_\mu)/N_\mu$, $N_\mu$
is the size of the lattice in the direction $\mu$, the integers
$0\le\ell_\mu<N_\mu$, and $\zeta_\mu=0$ for periodic boundary conditions
and 1 for anti-periodic.  Periodic or anti-periodic boundary conditions
imply that the independent modes are those with $\ell_\mu$ inside the
Brillouin hypercube whose body diagonal, BD, joins the corners (0,0,0,0)
and ($N_x/2,N_y/2,N_z/2,N_t/2$).

We used this power spectrum to find how smearing affects the UV and
IR modes.  We separated the IR and UV using hyperplanes perpendicular
to BD. All modes within the Brillouin zone closer to the origin
than a hyperplane $\sigma_\IR$ were called IR modes; conversely all
modes within the Brillouin zone closer to the far corner than the plane
$\sigma_\UV$ were called UV modes. Everything else was a generic mode--
neither IR, nor UV.  We defined the suppression of power in the IR and
UV as a function of $\epsilon$
\beq
   Q_\UV=\frac{E_\UV(\epsilon)}{E_\UV(0)}, \qquad{\rm and}\qquad
   Q_\IR=\frac{E_\IR(\epsilon)}{E_\IR(0)},
\eeq{qx}
where $E_\UV(\epsilon)$ is the power summed over all modes in the UV for
a fixed value of $\epsilon$, and $E_\IR(\epsilon)$ is a similar quantity
obtained by summing over all modes in the IR.  The definitions of IR and
UV are arbitrary, so one needs to check whether the results are sensitive
to this definition.  We placed the planes $\sigma_\IR$ and $\sigma_\UV$
at a fraction $d$ of the length of the diagonal (with $0<d<0.5$, so that
no mode is simultaneously in the IR and UV) from the nearest corner,
and varied $d$. We observed that results were insensitive to $d$.

We investigated $Q$ numerically with thermalized configurations at $T=0$
using $\beta=5.2746$ and $am=0.015$. Periodic boundary conditions were
used so that all $\zeta_\mu=0$.  The variation of $Q_X$ with $\epsilon$
is shown in \fgn{Q}. One sees that the slope of the curve for $Q_\UV$
always starts off larger than that for $Q_\IR$. Also, the slope of
the latter seems to be close to zero. This shows that smearing can be
used to modify the UV without modifying the IR.  One can use this to
seek an optimum value of $\epsilon$, such that $Q_\UV$ is as small as
possible. In a simulation with dynamical smeared quarks one would have
to do this without making a significant change in $Q_\IR$. In this study
the smearing is quenched; the set of gauge configurations is not changed
by smearing, only valence fermions are affected by smearing. So in this
context we are free to drop the condition on $Q_\IR$.

We also investigated the quark mass and lattice spacing dependence
of $Q_\IR$ and $Q_\UV$ by studying thermalized configurations at
$T=0$ using $\beta=5.2875$ and $am=0.025$ as well as $\beta=5.53$ and
$am=0.0125$. The first set has almost the same lattice spacing as the
one with $\beta=5.2746$, but has a somewhat different pion mass. The
last two sets have the same pion mass but lattice spacings which differ
by a factor of two. We show the results in \fgn{Q}. As can be seen very
clearly, there is a change in the overall suppression of power in the
IR and UV, but the change in the optimum $\epsilon$ is not large even
when the lattice spacing is halved. The optimum values of $\epsilon$
move down slightly. This movement is compatible with the intuition that
finer lattices require less improvement.

\bet
\begin{center}
\begin{tabular}{|c|ccc|ccc|}
\hline
Scheme 
 & \multicolumn{3}{c|}{$\beta=5.2875$, $am=0.025$}
 & \multicolumn{3}{c|}{$\beta=5.53$, $am=0.0125$} \\
\cline{2-7}
 & $Q_{UV}$ & $N_{CG}$ & $\lambda_{max}$ 
 & $Q_{UV}$ & $N_{CG}$ & $\lambda_{max}$ \\
\hline
APE   & 0.71 & 0.65 & 0.62 & 0.70 & 0.65 & 0.60 \\
HYP   & 0.65 & 0.60 & 0.56 & 0.65 & 0.55 & 0.55 \\
Stout & 0.19 & 0.15 & 0.16 & 0.18 & 0.15 & 0.14 \\
HEX   & 0.20 & 0.15 & 0.17 & 0.17 & 0.15 & 0.14 \\
\hline
\end{tabular}
\end{center}
\caption{The best $\epsilon$ for two different $a$, the second being half
 of the first, evaluated in different schemes and by different optimization
 criteria. The optimum parameter value in each scheme is nearly independent
 of $a$.}
\eet{opteps}

Interestingly, conjugate gradient inversion is also optimized at similar
values of $\epsilon$ \cite{condit}. In \fgn{cgkno} we show the number
of conjugate gradient iterations required to invert a smeared staggered
Dirac operator, $N_{CG}$, in a representative configuration drawn from
thermalized ensembles. These results were obtained with a CG stopping
criterion that the norm of the residual is less than $10^{-5}\sqrt
V$. Note that the performance of the APE and HYP smeared operators
are very similar to each other, just as before. The behaviour of the
Stout and HEX smearing are also similar, but quite different from the
previous pair. Again, the lattice spacing and pion mass seems to make
little difference to the optimization.

Using the smeared staggered Dirac operator, $D$, we found the minimum and
maximum eigenvalues of $D^\dag D$: $\lambda_{min}$ and $\lambda_{max}$. We
defined the condition number $\kappa(\epsilon) = \lambda_{max}(\epsilon)
/ \lambda_{min}(\epsilon)$. One expects that the number of CG iterations
is closely related to $\kappa(\epsilon)$, as indeed it is seen to be (see
\fgn{cgkno}). We found that $\lambda_{min}(\epsilon)$ is independent of
$\epsilon$ to better than 1\%, as expected, so $\lambda_{\min}(\epsilon)
/ \lambda_{min}(0)$ is flat. The dependence of $\kappa$ on $\epsilon$
is essentially due to the variation of $\lambda_{max}(\epsilon)$.

We can use any of these criteria, namely, the minimization of $Q_{UV}$,
$N_{CG}$, or $\lambda_{max}$, to choose the best value of $\epsilon$. The
results are shown in \tbn{opteps} in different smearing schemes at
different lattice spacings. There is a marginal decrease in the best
smearing parameter in each scheme with decrease in lattice spacing. We
see that there is reasonable agreement between the best values obtained
through the three methods. Given this, we choose to work with the values
$\epsilon=0.6$ for APE and HYP, and with $\epsilon=0.15$ for HEX and
$\epsilon=0.1$ for Stout.

\subsection{Smeared quarks and chirally symmetric correlators}\label{sec:corr}

\bef[thb]
\begin{center}
\includegraphics[scale=0.65]{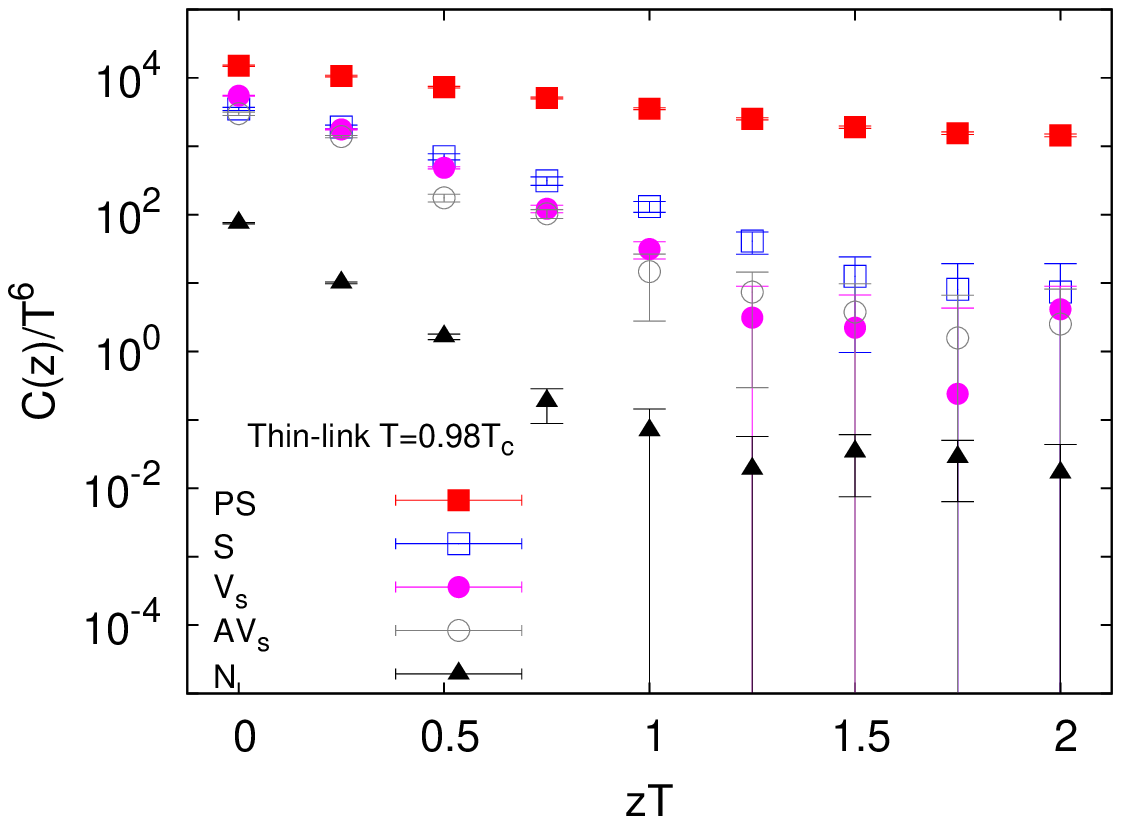}
\includegraphics[scale=0.65]{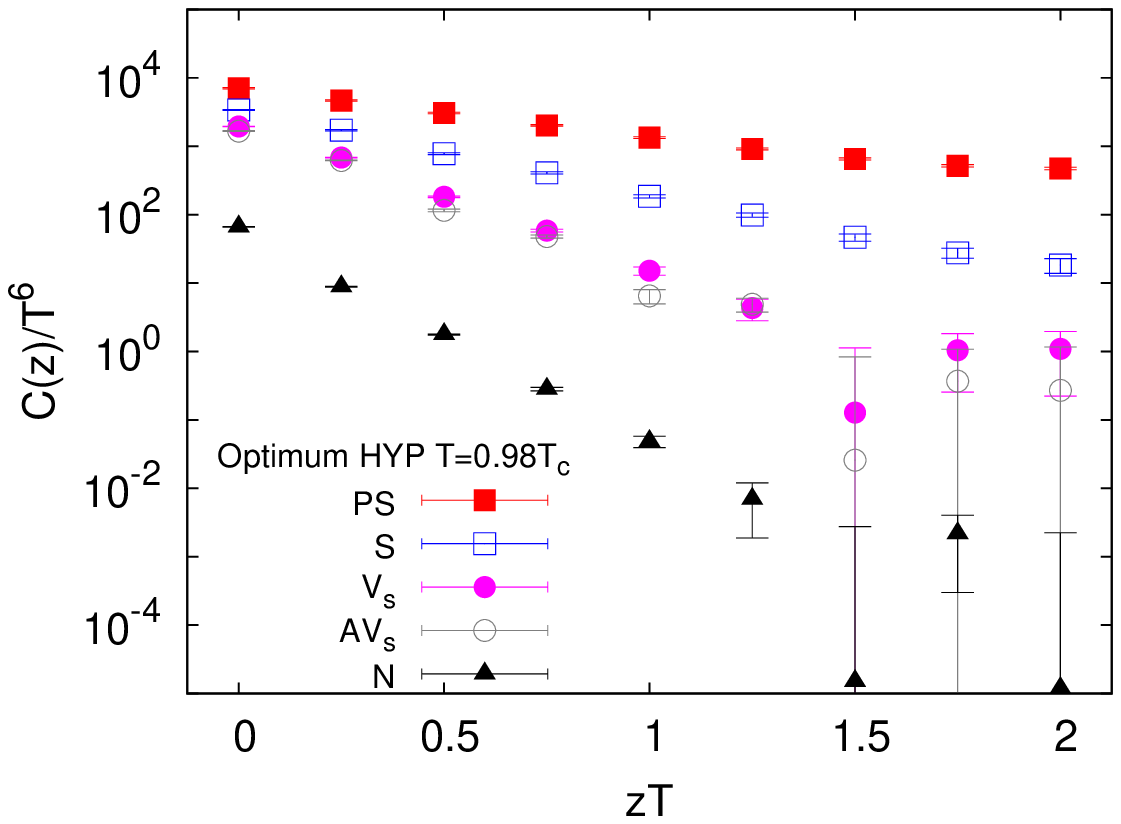}
\includegraphics[scale=0.65]{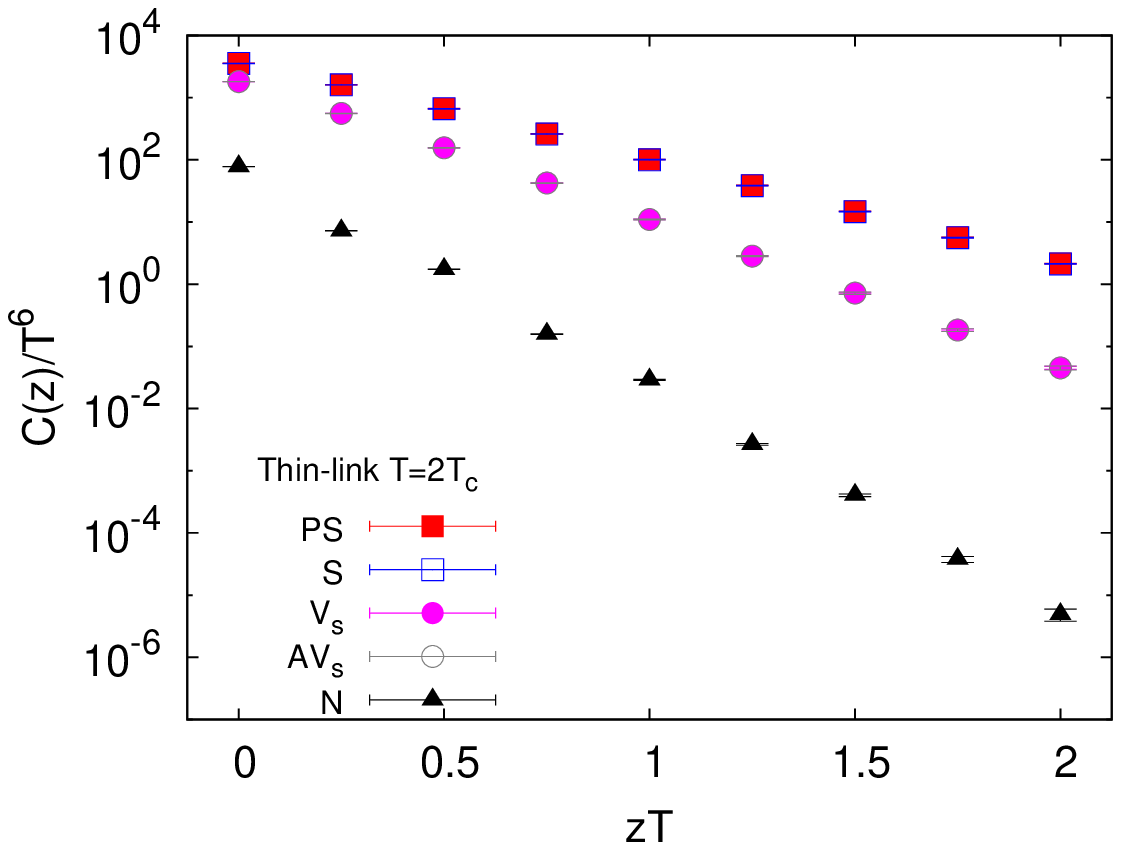}
\includegraphics[scale=0.65]{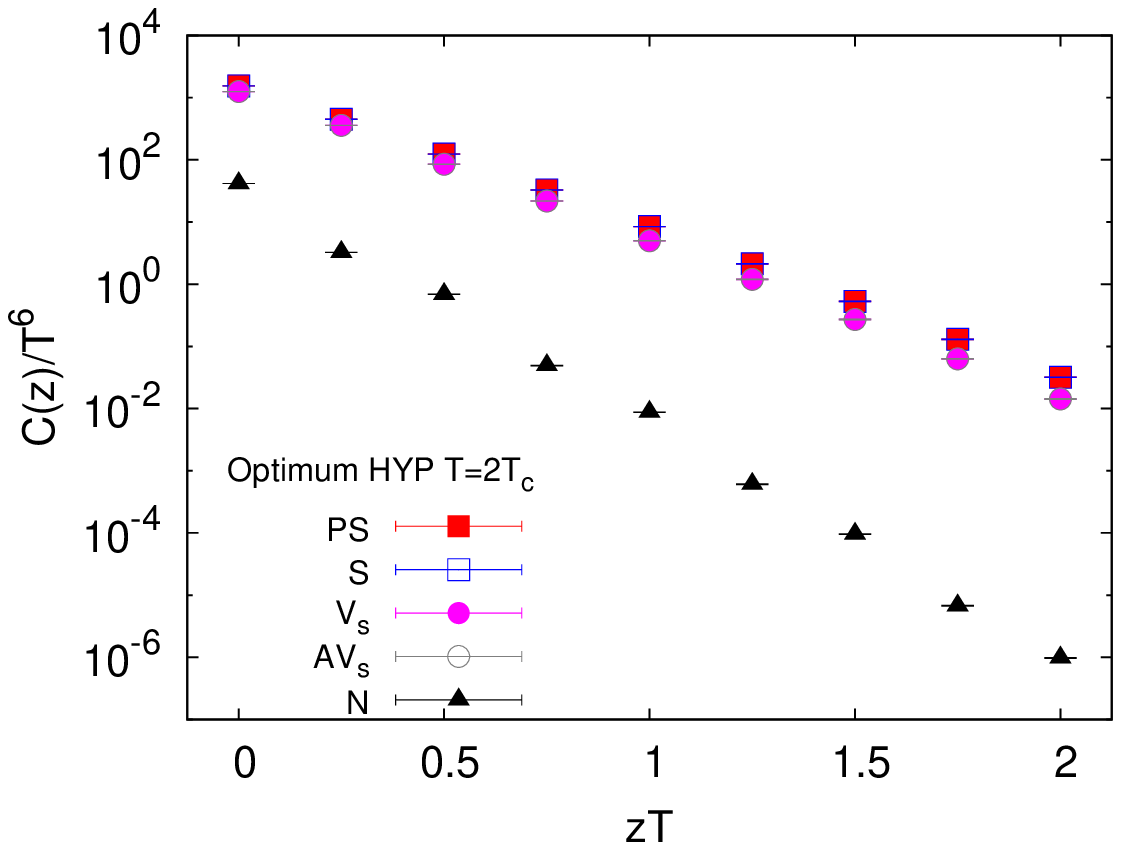}
\end{center}
\caption{Screening correlators from data set N above and below $T_c$. The
 signs of chiral symmetry restoration are clear with either thin-link
 or improved valence quarks in the form of pairwise degeneracies of
 correlators above $T_c$. However, improved correlators show even higher
 degeneracies at high temperature.  Similar results are obtained for
 data sets O and P.}
\eef{corlohi}

\bef[thb]           
\begin{center}
\includegraphics[scale=0.65]{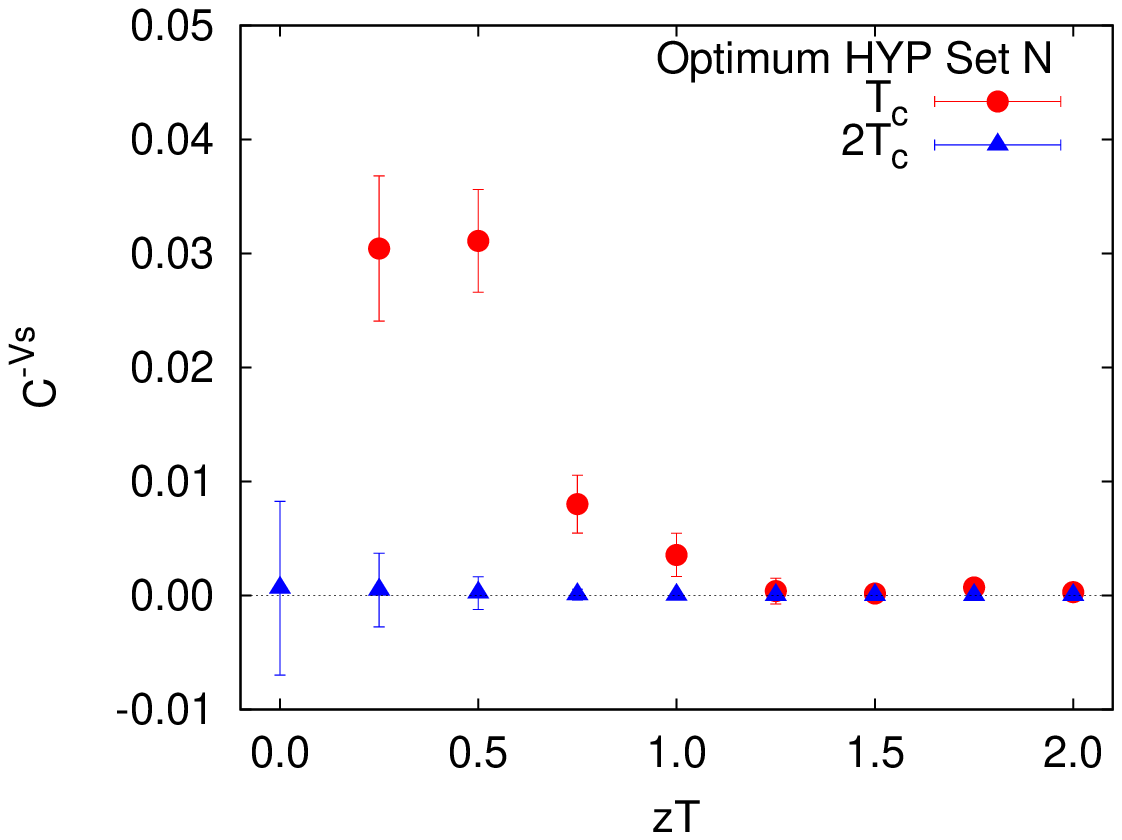}
\includegraphics[scale=0.65]{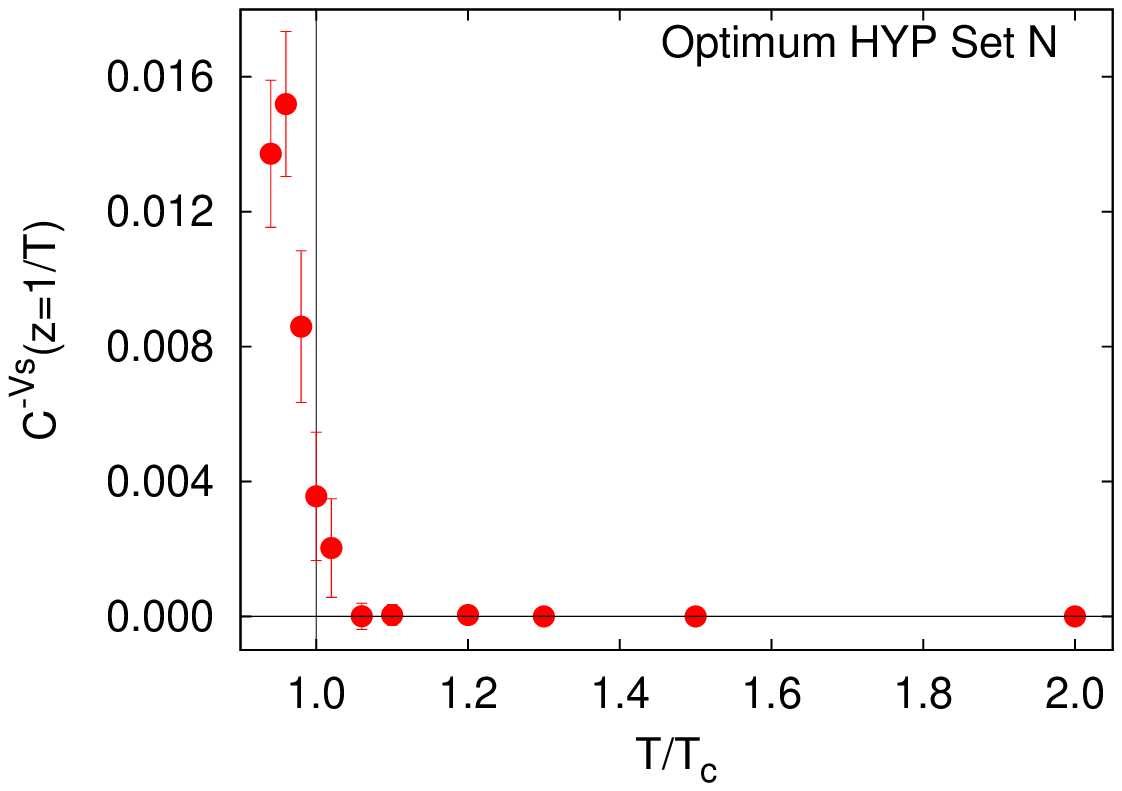}
\end{center}
\caption{The correlator $\prm{V_s}$ shows interesting short distance
 ($z\le1/T$) spatial structure slightly above $T_c$ (left). This effect
 barely persists into the hot phase (right). The results are displayed
 for set N. Sets O and P show similar behaviour.}
\eef{negchiproj}

\bef[bht]
\begin{center}
\includegraphics[scale=0.65]{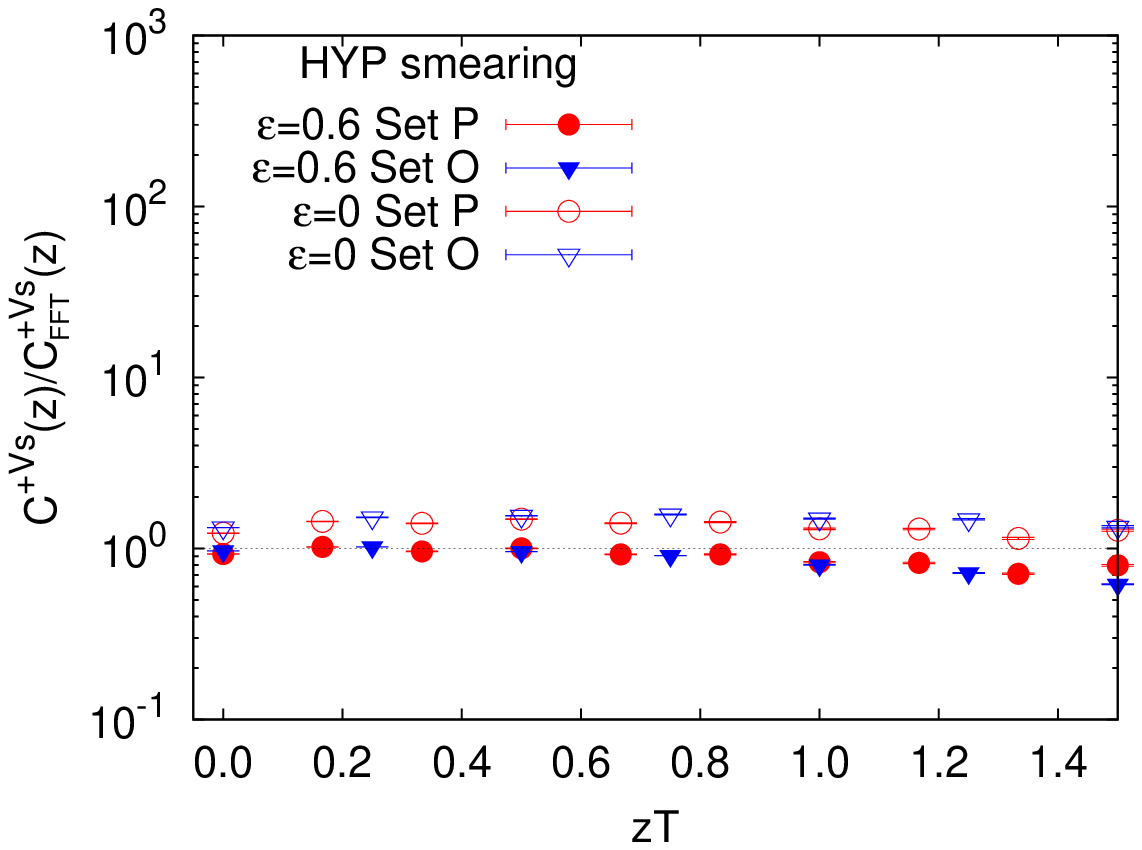}
\includegraphics[scale=0.65]{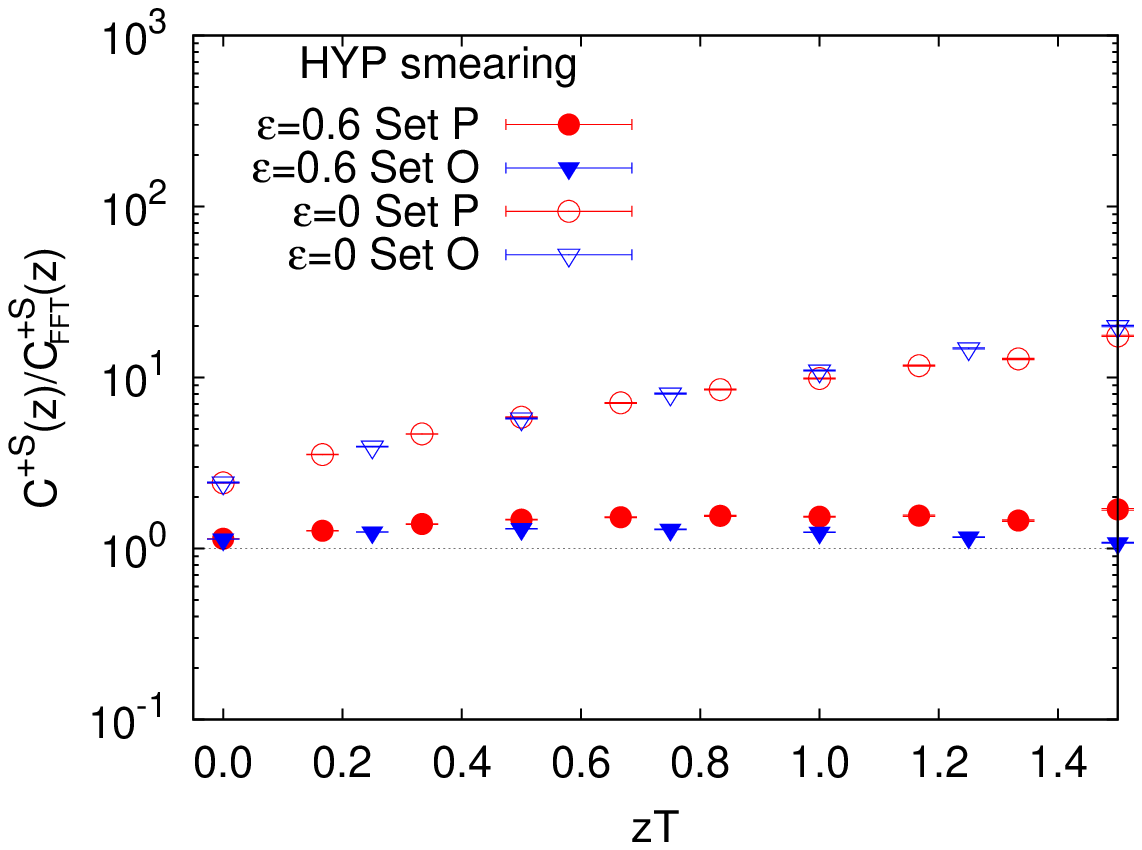}
\end{center}
\caption{The ratio of chiral projections $\prp{V_s}$ (left)
 and $\prp S$ (right) at $2T_c$ to the respective FFT prediction with
 data sets O and P. The smeared correlators come close to FFT in both
 cases, whereas the unsmeared $\prp S$ is quite different.}
\eef{chiralproj}

\bef[bht]
\begin{center}
\includegraphics[scale=0.8]{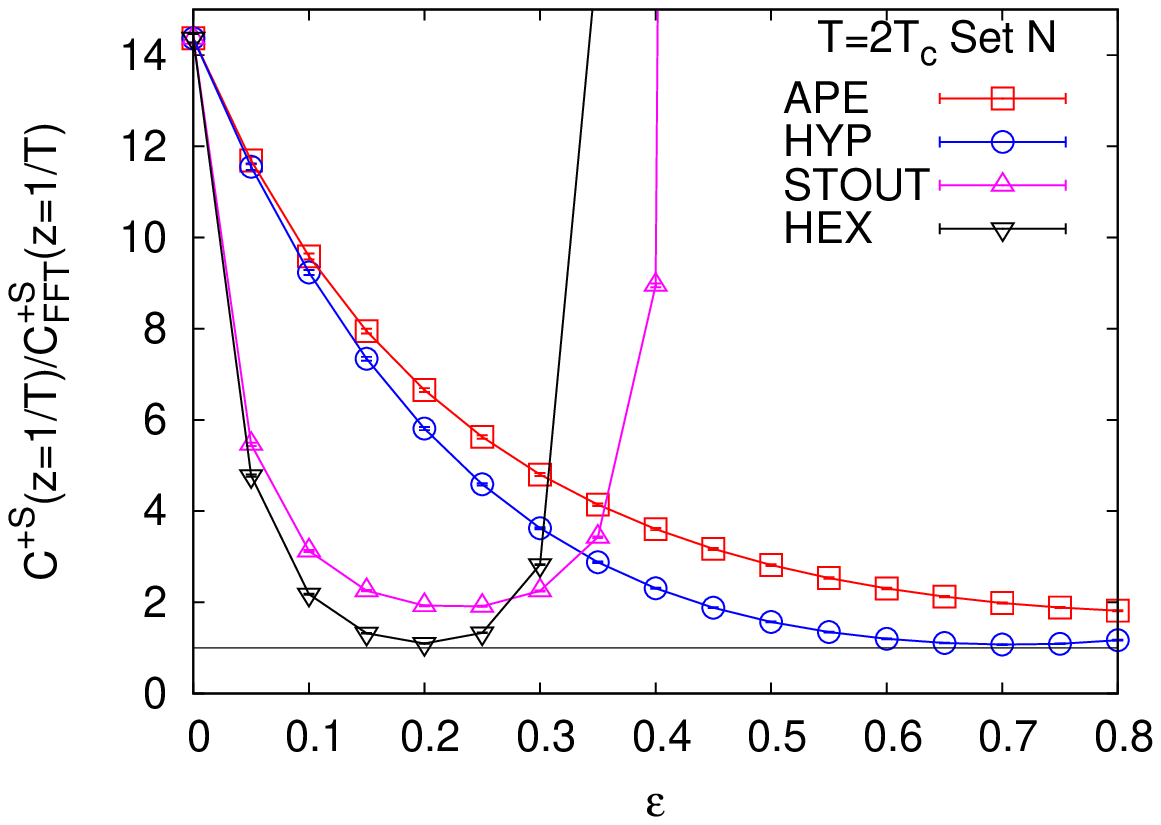}
\end{center}
\caption{The correlator $\prp S(z=1/T)$ at $2T_c$ from set N, normalized by
 its value in FFT, as a function of the smearing parameter $\epsilon$ in
 various smearing schemes.}
\eef{smearcor}

Chiral symmetry restoration in the high temperature phase of QCD is
easily seen in hadronic correlation functions. Below $T_c$ the local
meson correlators: S, PS, V, and AV are quite distinct, but above $T_c$
they collapse into one (see \fgn{corlohi}).  A pairwise degeneracy of the
S/PS and V/AV shows chiral symmetry restoration, and has been demonstrated
earlier as well with thin-link staggered valence quarks. However, the
near-degeneracy of the two pairs at high temperature, visible only after
smearing, is a new observation. This occurs in all the data sets: N, O,
and P.

Pairwise degeneracy arising from chiral symmetry restoration is most
easily seen in the vanishing of $\prm S$, $\prm{V_t}$ and $\prm{V_s}$
at high temperature \cite{quasi}. On examining these combinations, it
turns out that the degeneracy for $T>T_c$ becomes clearer with smearing.
For example, $\prm{V_t}(z=1/T)$ is $(6\pm8)\times10^{-3}$ at $T_c$ with
thin-link valence quarks, but becomes $(0\pm2)\times10^{-3}$ when optimal
HYP smeared valence quarks are used. The improvement is most remarkable in
the S/PS sector, where we found $\prm S(z=1/T)=-3.3\pm0.1$ at $T_c$ using
thin-link valence, but $-0.57\pm0.04$ using optimal HYP smeared valence.
At larger $T$ all the negative chiral projections vanished. It was
seen earlier \cite{quasi} that $\prm{V_s}$ for $T\ge T_c$ vanished when
$z>1/T$, but remained non-zero at short distances. In \fgn{negchiproj}
we show this effect at $T_c$ and also, that it vanishes at $2T_c$. A more
detailed view of the temperature dependence is exhibited by showing how
$\prm{V_s}(z=1/T)$ changes with $T$.  Below $T_c$ the correlator does
not vanish, but the spatial structure seems to have entirely disappeared
for $T>1.05T_c$. This gives one definition of the width of the chiral
crossover; it is larger than the one implied by $\Delta\beta_c$ (see
Appendix \ref{sec:tc}).

$\prp{V_s}$ is close to the FFT prediction with either thin-link or
smeared valence quarks.  With thin-link staggered valence quarks, we see
that $\prp S$ is different from FFT, as previously observed. However,
on smearing, they become compatible with FFT (see \fgn{chiralproj}).
This is a more detailed understanding of why the meson screening
correlators are nearly degenerate in \fgn{corlohi}.

In \fgn{smearcor} we show that the correlator $\prp S(z)$ approaches
FFT as the parameter $\epsilon$ is tuned to the optimum in each of the
smearing schemes, approaching closest to FFT at the optimum.  We used the
distance $z=1/T$ in this demonstration because it is neither in the far IR
nor in the UV. The optimum HYP and HEX schemes bring the correlator closer
to FFT than the APE and Stout smearing schemes, although the latter also
come very close to FFT. Next we explore this difference between schemes.

\subsection{Smearing, taste symmetry and screening masses}\label{sec:scr}

\bef[bth]
\begin{center}
\includegraphics[scale=0.7]{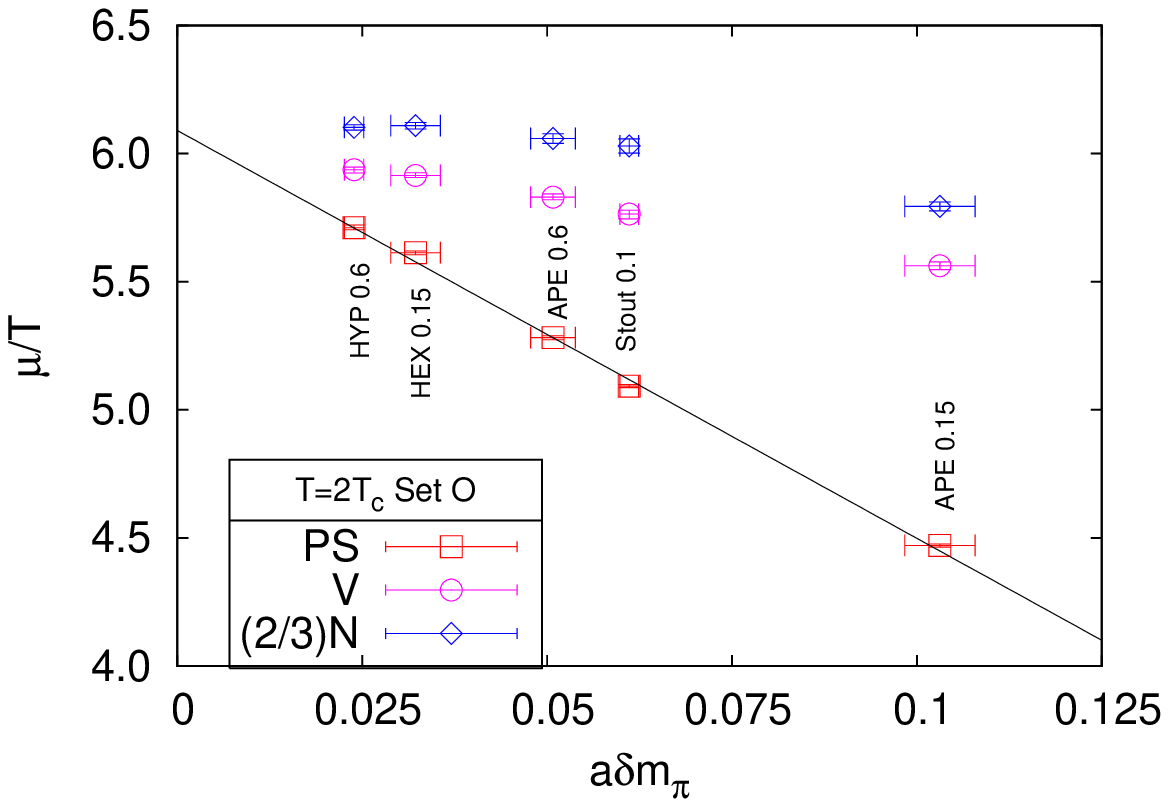}
\includegraphics[scale=0.7]{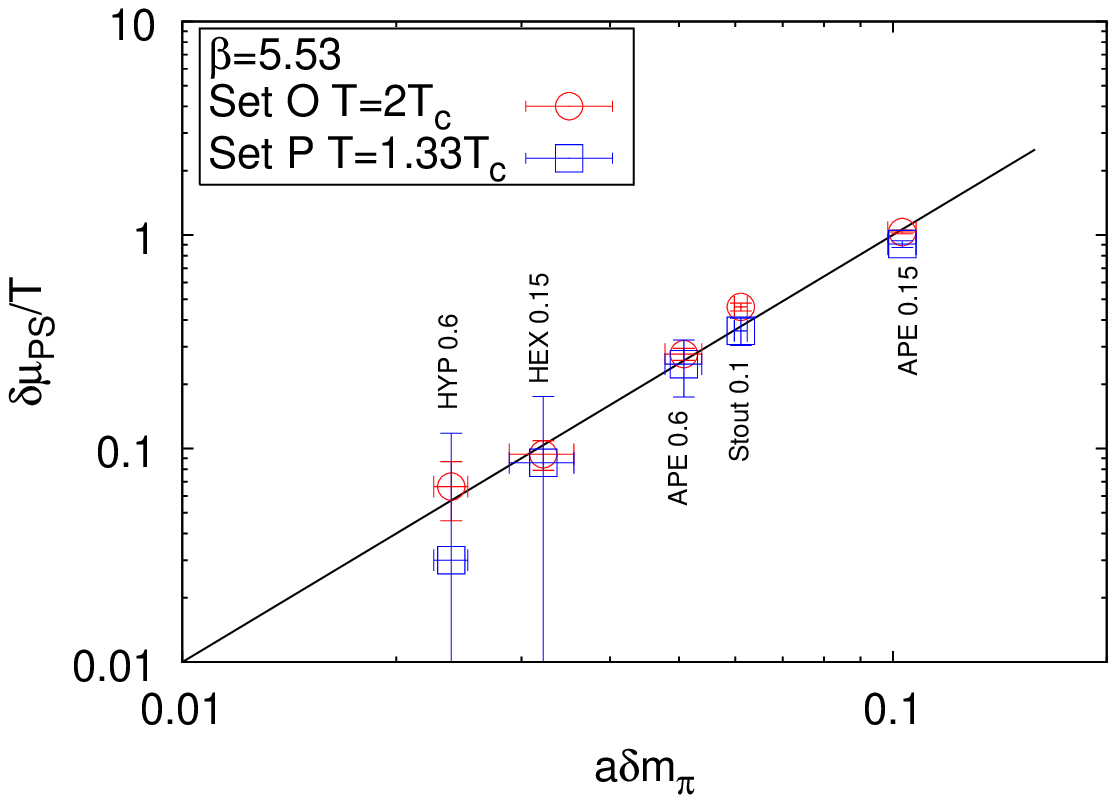}
\end{center} 
\caption{The first panel shows screening masses of the local S/PS,
 V/AV and N at $2T_c$ as functions of the pion taste splitting $a\delta
 m_\pi$ at $2T_c$ in data set O. The screening mass of the nucleon has
 been multiplied by $2/3$ in order to compress the vertical scale. Each
 set of screening masses varies linearly with a measure of the pion taste
 splitting, $a\delta m_\pi$.  The second panel plots $a\delta m_\pi$
 at $T=0$ against the splitting of the corresponding screening masses,
 $a\delta\mu_\PS$ at two different temperatures, but at the same lattice
 spacing; the line $y=100x^2$, is superposed to indicate the slope.}
\eef{splitscreen}

On examining screening masses, we found that they depend on the smearing
parameter $\epsilon$ essentially only through the taste symmetry breaking
measure
\beq
   \delta m_\pi=m_{\gamma_5\gamma_i}-m_{\gamma_5},
\eeq{split}
where the subscripts on the right denote the pion taste structure.
The $\gamma_5$ taste is the Goldstone pion. We chose the partner with
taste structure $\gamma_5\gamma_i$ as an indicator of taste splitting
since it turned out to be relatively easily measured. \fgn{splitscreen}
shows the nearly linear dependence of $\mu/T$ on $\delta m_\pi a$. The
figure shows the clear superiority of the HEX scheme over Stout. Using
the scaling shown in \fgn{splitscreen}, one could extrapolate screening
masses to the limit $\delta m_\pi\to0$. However, this is premature, since
it involves an extrapolation to suboptimal values of $\epsilon$. The S/PS
screening masses obtained using local operators with dynamical p4 quarks
at a comparable temperature turns out to be around $4.8T$ \cite{edwin}.

More information can be extracted from the taste-splitting of the screening
masses at finite $T$,
\beq
   \delta\mu_\PS = \mu_{\gamma_5\gamma_i}-\mu_{\gamma_5}.
\eeq{Tsplit}
The only previous study of this kind was reported in \cite{edwin}. In
\fgn{splitscreen} we show $\delta\mu_\PS$ as a function of $a\delta m_\pi$.
In making this comparison we held the lattice spacing fixed, with one
set of measurements at $T=0$, one at $T=2T_c$ in set O and a third at
$T=1.33T_c$ in set P. We find $\delta\mu_\PS\propto T(a\delta m_\pi)^2$
over the range of values we obtained. This removes the ambiguity
remarked upon in \cite{edwin}.

One can argue for this on general grounds.  A hadron mass, $M$, can be
written as $Ma=f(a\Lambda_\MSbar,ma,\epsilon)$, where we treat $\epsilon$
as a generic label for all the parameters which control smearing. A
screening mass, $\mu$, can be written as $\mu/T = g(a\Lambda_\MSbar, ma,
\epsilon, N_t)$, since $N_t=1/(aT)$, or as $a\mu = g'(a\Lambda_\MSbar,
ma, \epsilon, N_t)$.  For data taken at fixed cutoff, $a\Lambda_\MSbar$,
we need not show this parameter explicitly.  Although we work at fixed
$ma$, it is profitable to consider the dependence on this variable. A
series expansion in $ma$ near the chiral limit would yield different
dependence for the Goldstone pion mass
\beq
   am_{\gamma_5} = \alpha_1\sqrt{ma} + {\cal O}\left(ma\right)^{3/2},
    \qquad{\rm and}\qquad
   aM = \gamma_0 + \gamma_1 ma + {\cal O}\left(ma\right)^2,
\eeq{Aseries}
where $M$ is any other mass scale, and the coefficients depend on
$\epsilon$. As a result, $a\delta m_\pi = \gamma_0 - \alpha_1\sqrt{ma}
+ \gamma_1ma$ and $\delta\mu_\PS/T = \gamma_0' + \gamma_1'ma$.
If taste symmetry were recovered in the chiral limit by tuning
$\epsilon$, then one might argue that $\gamma_0=\gamma_0'=0$ and hence
$a\delta\mu_\PS\propto(a\delta m_\pi)^2$.  This would also mean that
all pion tastes are forced to be Goldstones, with an expansion starting
at order $\sqrt{ma}$. Chiral logarithms, which we have neglected here,
could become important at smaller masses and spoil this scaling.

Even if smearing achieves a more limited goal of significantly decreasing
$\delta m_\pi$ at finite $ma$ without actually recovering taste symmetry
completely, one might still recover quadratic scaling. All that is needed
is that $\gamma_0$ and $\gamma_0'$ become much smaller than the actual
$T=0$ taste splitting in the problem. In general one would have
\beq
   \delta\mu_\PS/T -\gamma_0' \propto (a\delta m_\pi-\gamma_0)^2.
\eeq{Ascaling} 
The data in \fgn{splitscreen} shows that $\gamma_0'$ and $\gamma_0$,
are small compared to $a\delta m_\pi$. This quantifies how well
smearing works. The fact that it seems to work better at finite temperature
than at $T=0$ with fixed values of $a\Lambda_\MSbar$ and $ma$ possibly
indicates that the Dirac eigenvalue spectrum is simpler. We shall return
to this point later.

The main conclusion from these studies of smearing is the following.
Optimising the suppression of UV modes automatically improves taste
symmetry in the hadron spectrum at $T=0$. This leads to superlinear
improvement in taste symmetry in the hot phase of QCD. In order to gain
most from such an improvement, one should then choose the best possible
smearing scheme. With partial quenching, as here, this would mean working
with the optimized HYP scheme; with dynamical smeared quarks it would
mean working with the optimized HEX scheme.

\section{Results}\label{sec:results}

\subsection{Hot QCD: weak coupling and the Dirac spectrum}

\bef[tbh]
\begin{center}
\includegraphics[scale=0.7]{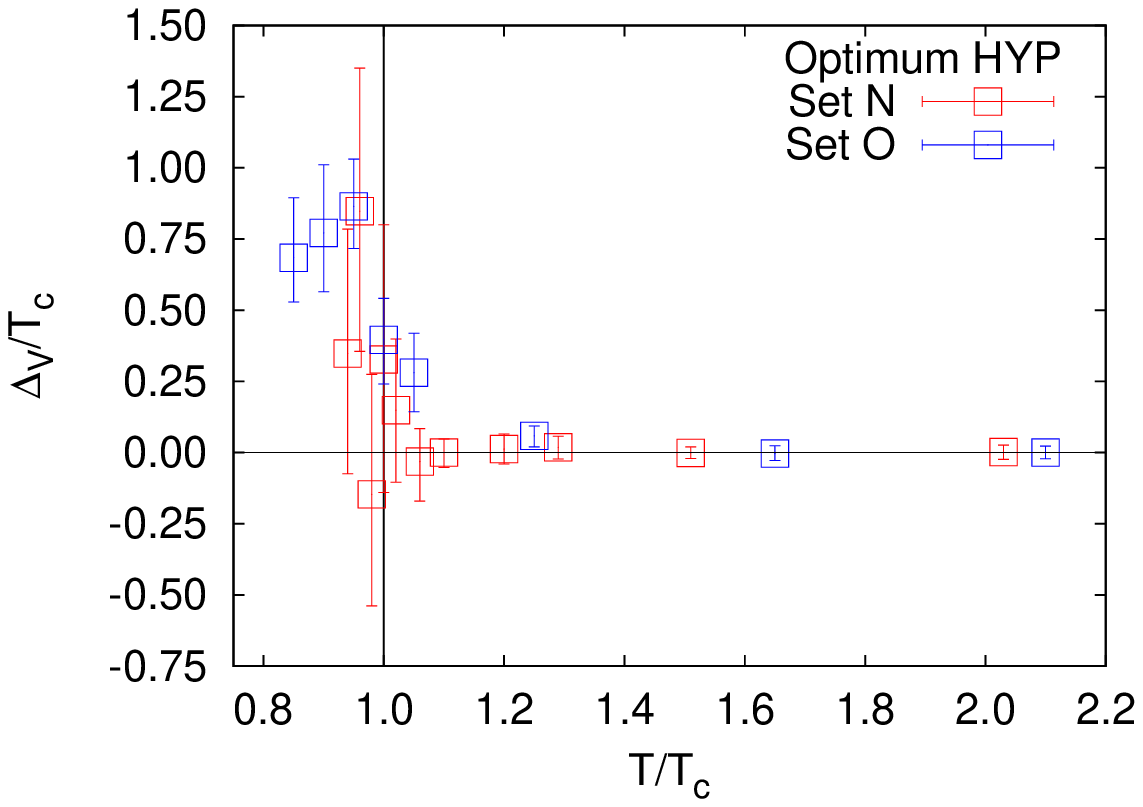}
\includegraphics[scale=0.7]{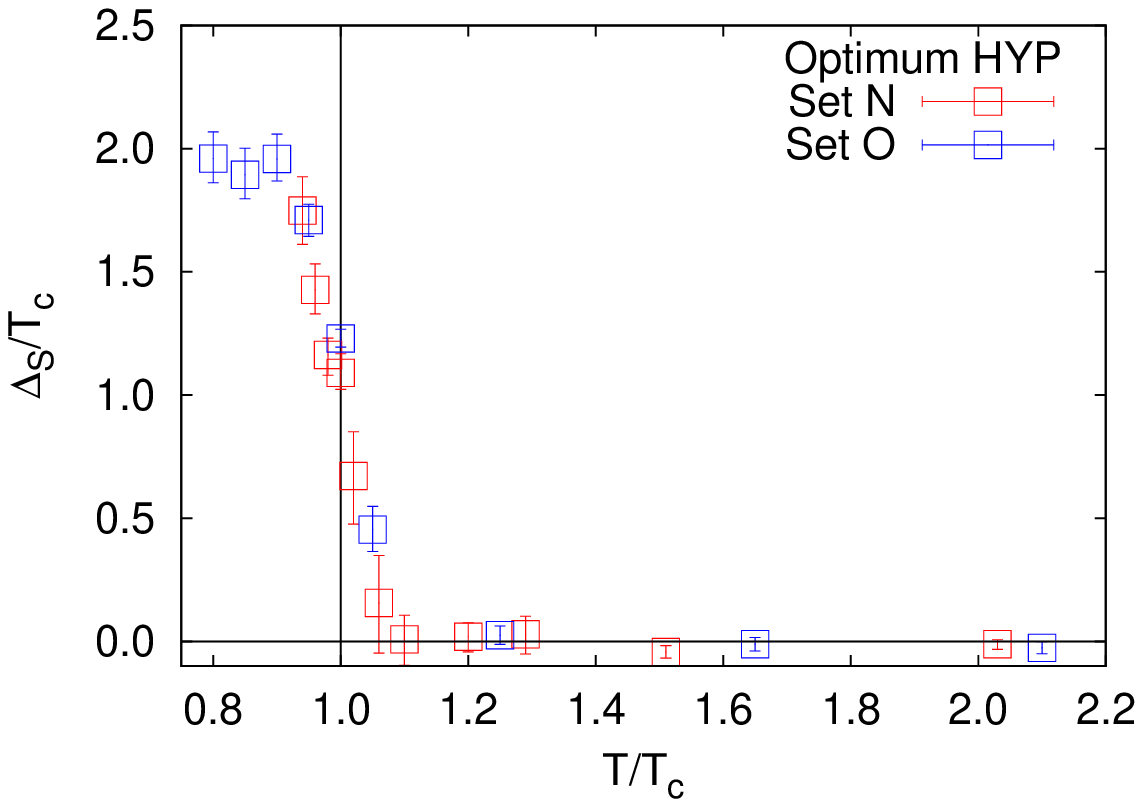}
\end{center} 
\caption{Variation of $\Delta_V$ (left) and $\Delta_S$ (right) with
 temperature.  $\Delta_V$ approaches 0 at the 95\% confidence level immediately
 above $T_c$, independent of smearing scheme and quark mass.}
\eef{delta}

\bef[tbh]
\begin{center}
\includegraphics[scale=0.75]{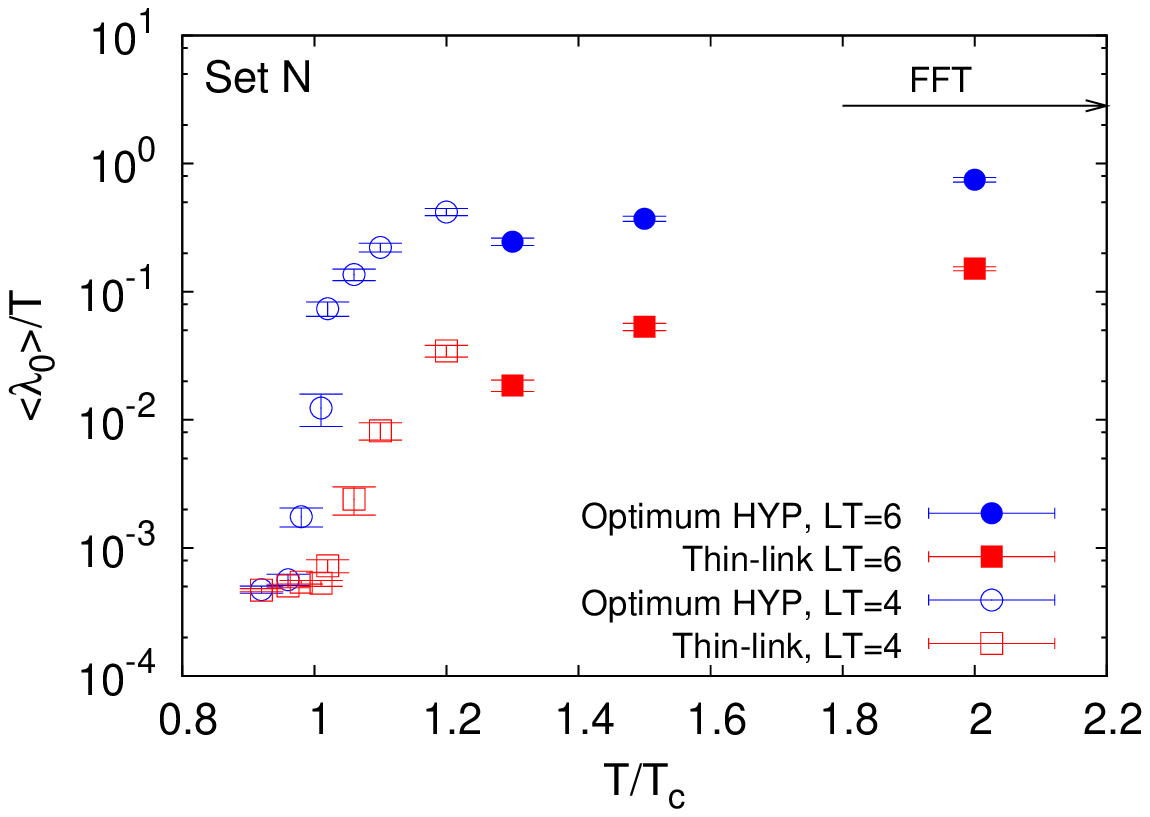}
\end{center} 
\caption{The ensemble averaged smallest eigenvalue of the massless staggered
 Dirac operator, $\langle\lambda_0\rangle$, for set N. With optimal HYP
 smearing, the eigenvalue rises by two orders of magnitude in a narrow range
 above $T_c$. For the thin-link Dirac operator, the rise is much slower.}
\eef{axial}

\bef[p]
\begin{center}
\includegraphics[scale=0.80]{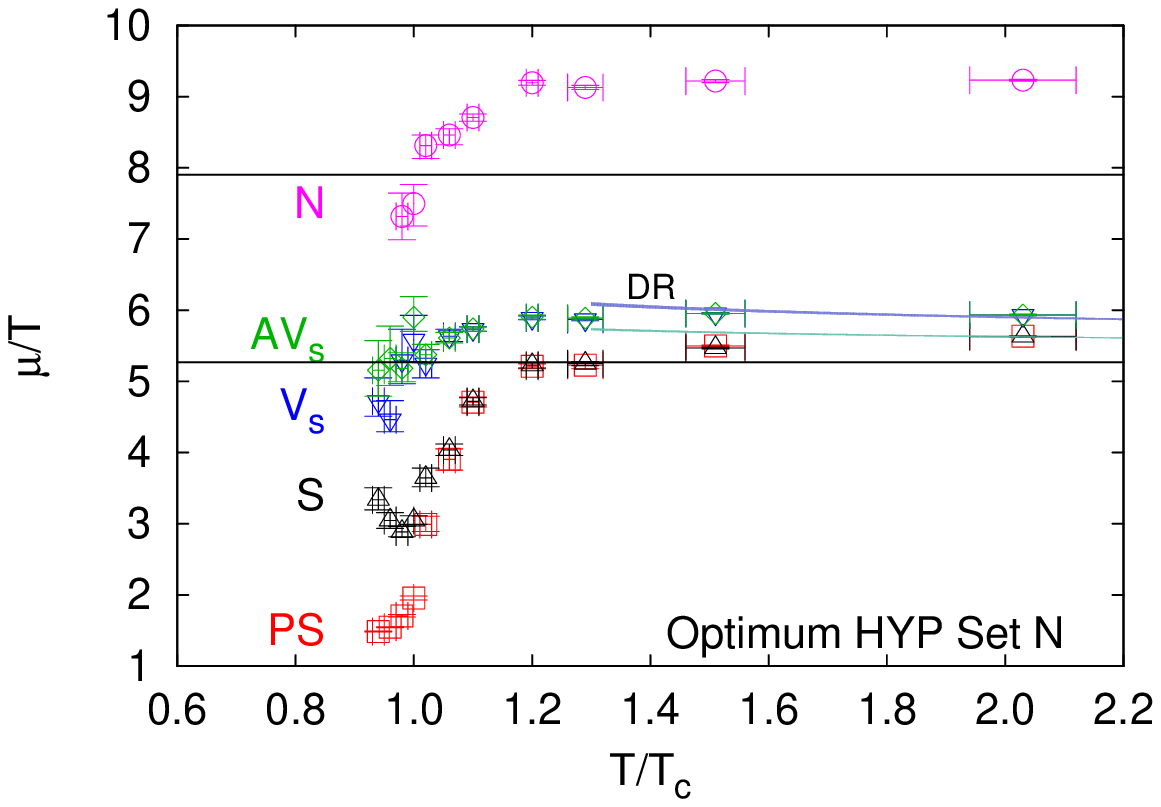}
\includegraphics[scale=0.80]{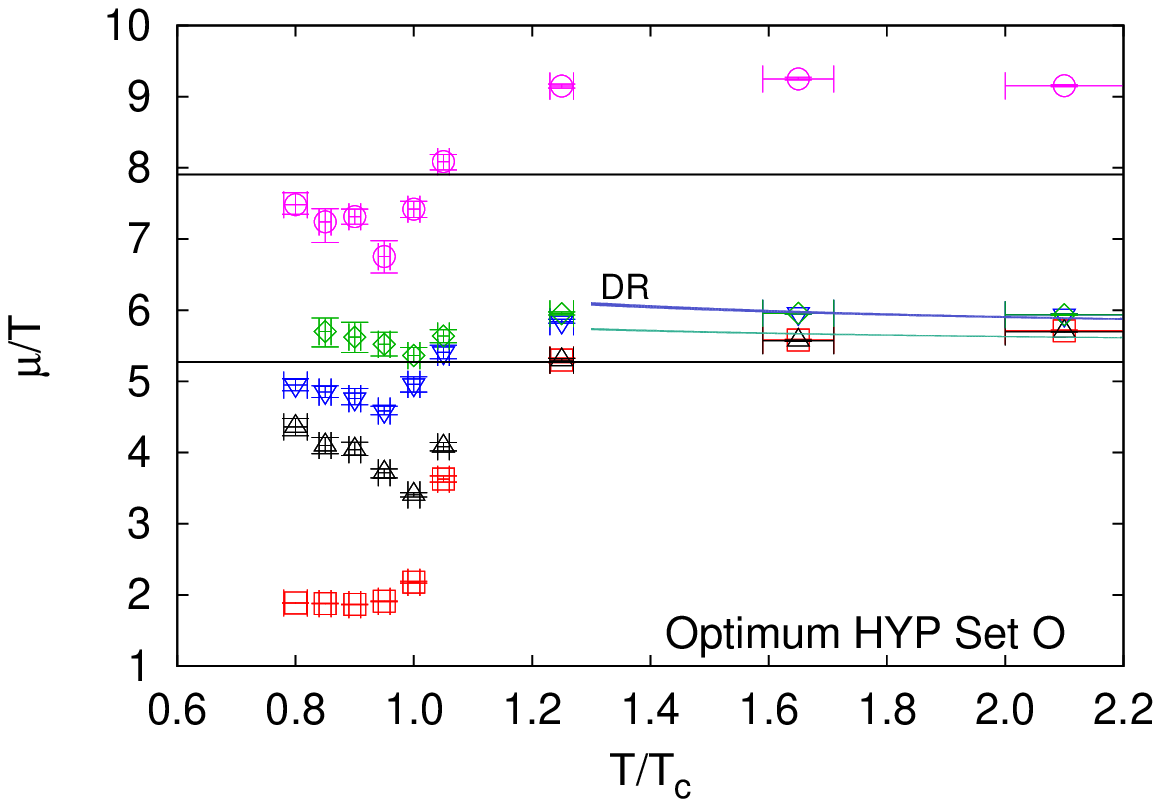}
\includegraphics[scale=0.80]{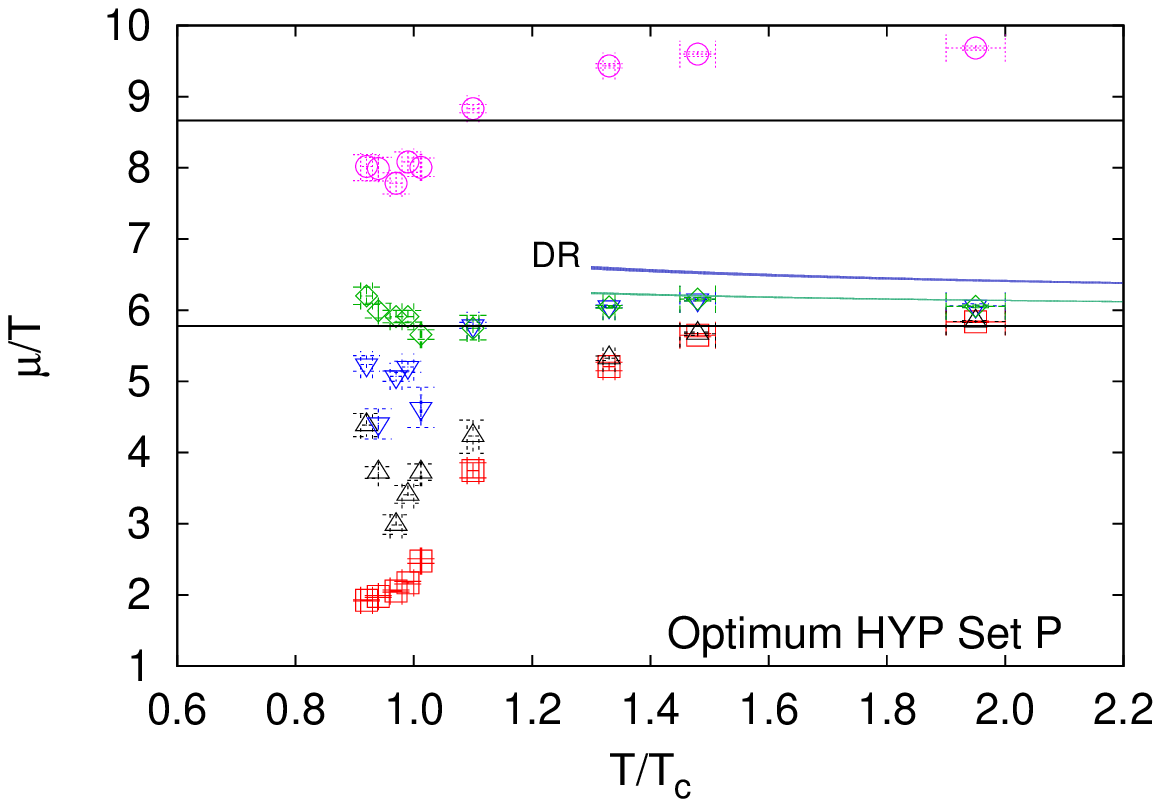}
\end{center} 
\caption{Hadron screening masses for the data sets N, O, and P using
 optimum HYP smeared correlators. The horizontal lines are the free
 theory screening mass for the nucleon and the mesons respectively:
 for set N and O they are 5.27 for mesons and 7.90 for baryons, for
 set P they are 5.78 for mesons and 8.67 for baryons. DR denotes the
 weak-coupling prediction of \cite{dr}; the unlabelled line immediately
 below this shows the prediction of \cite{htl}.}
\eef{spectrum}

We found that the mass-splitting between chiral partners changes rapidly
in the low-temperature phase and vanishes fairly close to $T_c$ in the hot
phase. In \fgn{delta} we show that $\Delta_S$ and $\Delta_V$ both vanish
at $T=1.05T_c$.  Also, a comparison of sets N and O show very little
quark mass dependence at about the same lattice spacing. These results
are in contrast to the observations in \cite{quasi,cheng} that $\Delta_S$
remains significantly non-zero up to a temperature significantly higher
than $T_c$.  The change from the old results \cite{quasi} using the same
data set P, confirms that the improvement is due to smearing.

The rapid approach to behaviour similar to weak-coupling theory has
implications for the spectrum of the staggered Dirac operator. The
vanishing of the pion mass in the chiral limit at $T=0$ is related to a
finite density of the Dirac eigenvalues near zero. It was shown earlier
in a study of set O with thin-link quarks that a gap developed in the
massless staggered eigenvalue spectrum a little above $T_c$, and that
the hot phase contained localized Dirac eigenvectors \cite{lacaze}.

Here we studied the gap by measuring the smallest eigenvalue of the
massless staggered Dirac operator, $\lambda_0$. The ensemble average,
$\langle\lambda_0\rangle$, at a given temperature was generally
seen to be within a factor of four of the minimum over the ensemble.
In view of this, we report $\langle\lambda_0\rangle$. As can be seen in
\fgn{axial}, it climbs by two orders of magnitude between $T_c$ and
$1.06T_c$ for the smeared Dirac operator. For the thin-link operator,
$\langle\lambda_0\rangle$ rises at significantly higher temperature.

One sees some volume dependence in the result. This was studied
extensively in \cite{lacaze}, where it was found that the volume
dependence becomes negligible when the spatial size, $L$, is of the
order of $1/\langle\lambda_0\rangle$. For $LT=4$, this would be at
$\langle\lambda_0\rangle\simeq0.25$, which seems to happen at $1.5T_c$. In
future it would be interesting to study this volume dependence further.

It is also of interest to note that $a\langle\lambda_0\rangle$ becomes
comparable to $am$ at $T=T_c$ with optimum HYP smearing. Since this
happens for all the data sets, it accounts for the lack of quark mass
dependence seen in $\Delta_V$ and $\Delta_S$. With thin-link quarks this
crossing is delayed to $T/T_c\simeq1.5$, thus affecting all screening
phenomena.

In set N at $T=1.3T_c$ we spotted one configuration out of the 50 for
which $\lambda_0$ was two orders of magnitude below $\langle \lambda_0
\rangle$.  This implies the existence of a small fraction of atypical
configurations in the thermal ensemble. These would be interesting
in a study of axial U(1) symmetry at finite temperature, where such
atypical configurations have been linked to topological configurations by
observations with overlap \cite{oldlacaze} or HISQ quarks in \cite{ohno}.
However, that would require a much larger statistical sample, and is
therefore best left to the future.

\subsection{Comparison with weak-coupling theory}

Finally, the results for the screening masses as functions of $T$
are shown in \fgn{spectrum} for all three data sets with optimal HYP
smearing. Also shown are the values expected in FFT on lattices with
the same size. The analysis of correlation functions obtained with these
smeared valence quarks shows that the screening masses in all channels
approach FFT at high $T$. The most striking new feature of this data is
that this approach is from above, in conformity with the predictions of
\cite{dr}. Similar results are obtained with optimal HEX smeared
quarks.  We have shown earlier in \fgn{splitscreen} that there is a
remaining uncertainty of around 15\% in the determination of the the S/PS
screening mass. This comes from the residual taste symmetry breaking at
the best optimization of the screening parameters possible at this lattice
spacing. Reduction of this uncertainty requires going to finer lattices.

The weak coupling prediction for the meson-like screening masses is
\beq
   \mu = \mu_{FFT} + \frac43\alphas[1+2E_0]T.
\eeq{weak}
Here $\alphas$ is the 2-loop QCD coupling evaluated in the $\overline{MS}$
scheme at the scale $2\pi T$.  $E_0=0.3824$ for two flavours of quarks in
a dimensional reduction (DR) scheme \cite{dr}.  A hard thermal loop (HTL)
resummation which neglects soft gluon contributions to the vertex yields
$E_0=0$ \cite{htl}.  These weak-coupling predictions are also shown in
\fgn{spectrum}, with $\alphas$ determined using \cite{tscale}. As one
can see, both the weak-coupling predictions are close to the observed
screening masses.

\section{Conclusions}\label{sec:conclude}

Several properties of quarks at experimentally accessible temperatures
above $T_c$ seem to be explained in weak coupling QCD. However, one which
showed puzzling departures from weak-coupling predictions was screening
masses from hadronic excitations. In quenched computations it was seen
that the results depended strongly on the kind of valence quark used
\cite{scrnp}. With this clue in hand we performed computations with
dynamical staggered sea quarks and improved valence quarks in three
sets of computations, one new (set N, see \tbn{confnew}) and two older,
(sets O \cite{nt4} and P \cite{nt6}). Studies with staggered valence
quarks were reported earlier with set P \cite{quasi}.

A preliminary part of this work was the optimization of the valence
quarks.  We used four popular versions of fat-link staggered quarks.
We optimized the smearing parameter, $\epsilon$, in each case by
observing changes to the power spectrum of the plaquette (see \fgn{Q})
and the largest and smallest eigenvalues of the Dirac operator (see
\fgn{cgkno}). The optimum $\epsilon$ was chosen so that the UV was
suppressed as much as possible without changing the IR behaviour
in both cases.  This also improved the performance of the conjugate
gradient algorithm used for the inversion of the Dirac operator (see
\tbn{opteps}). Such a tuning was done at $T=0$. We found mild changes in
the tuning parameters as the lattice spacing was changed by a factor of 2.

Although the smearing parameter is optimized by requiring that the IR
components of fields do not change appreciably, it does affect the
long-distance properties of the theory, such as masses. We compared
different schemes through a measure of the recovery of staggered quark
taste symmetry in the spectrum of pions (see \fgn{splitscreen}). The
optimized HYP smearing works best, although optimized HEX smearing comes
a close second. This is pleasant, since dynamical simulations with HEX
smearing are easier than with HYP.

Smearing causes systematic changes in finite temperature properties
of interest. We found that the screening mass in the hot phase
increases systematically as taste symmetry breaking is reduced at
$T=0$ (see \fgn{splitscreen}).  Also, taste symmetry breaking in
the hot phase improves super-linearly with improvement at $T=0$ (see
\fgn{splitscreen}).  Since recovery of taste symmetry has been used
as the main indicator of the reduction of UV effects, it is natural
in this study to use optimized HYP smearing in order to best reduce
lattice artifacts.

On doing this we find that the screening correlator recovers the
degeneracies that a theory of weakly coupled fermions would predict
(see \fgn{corlohi}).  This happens very close to, and above, $T_c$
(see \fgn{negchiproj}).  The correlators themselves are also close to
the predictions of a free fermion field theory (see \fgn{chiralproj}).
Consistent with this, the screening masses at high temperature are found
to be close to weak-coupling theory (see \fgn{spectrum}).  A computation
in dimensional reduction \cite{dr} gives results which are slightly
different from a HTL computation neglecting soft-gluon effects on the
vertex \cite{htl}. The lattice computation is unable to distinguish
these as yet, but we may expect this to improve in the near future.

We also see that the smallest eigenvalue of the optimally HYP smeared
massless staggered Dirac operator shows a rapid jump from extremely
small values in the mean below $T_c$ to fairly large values above
(see \fgn{axial}). The behaviour of the thin-link staggered operator is
qualitatively similar, although quantitatively slower to change. Since
the smallest eigenvalue of the massless smeared operator is comparable to
the bare mass already at $T=T_c$, the limit of physical renormalized mass
becomes easy to take in the high temperature phase. There is evidence
for a very small fraction of completely atypical configurations in the
hot phase. A study of the topology of these gauge configurations lies
outside the scope of this paper.

The lattice computations described here were performed on the Cray
X1 of the ILGTI in TIFR. We thank Ajay Salve and Kapil Ghadiali for
technical assistance.  We also thank Debasish Banerjee, Saumen Datta,
Mikko Laine, Nilmani Mathur, Subroto Pal and Christian Schmidt for their
comments. Some of the configurations used in this study were generated
earlier for other studies by the ILGTI.

\appendix
\section{Determination of $\beta_c$}\label{sec:tc}

\bef
\begin{center}
\includegraphics[scale=0.7]{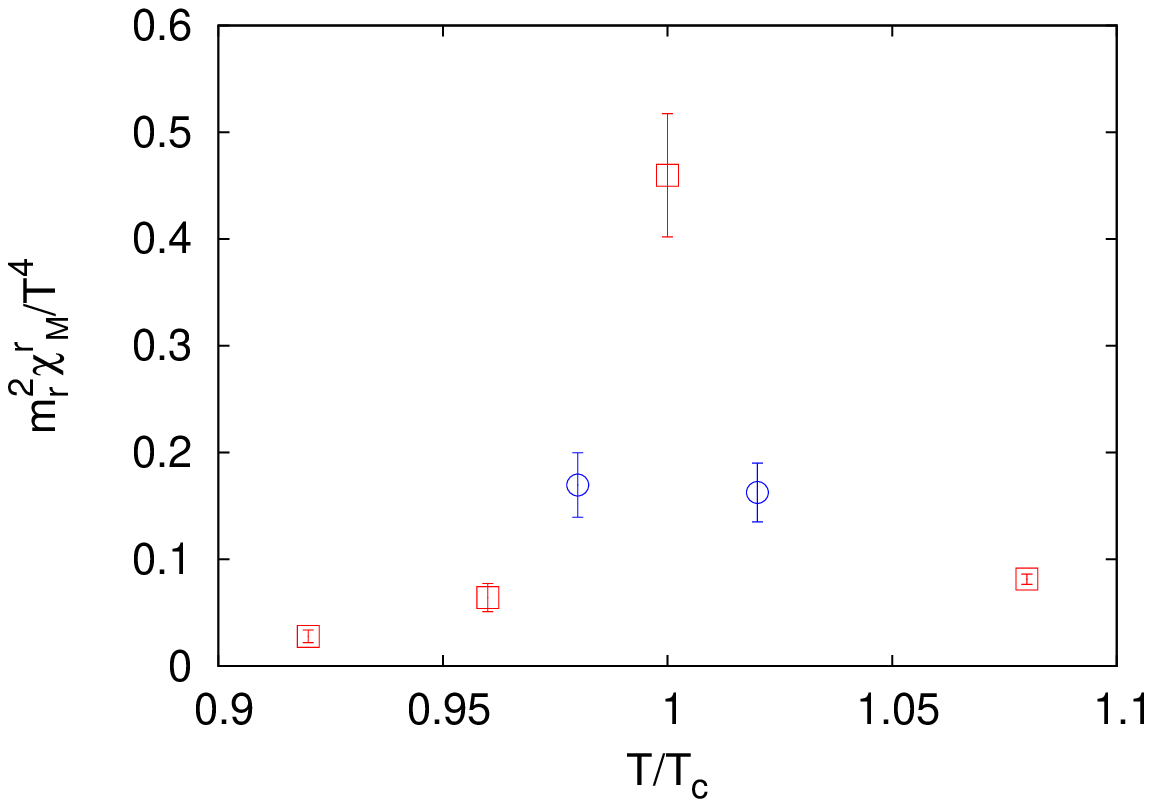}
\end{center}
\caption{Dependence of $m_r^2\chi_\M^r/T^4$ on $T/T_c$. The data points
 marked by boxes are measurements made on line of constant $m/T_c$, whereas
 those marked by circles are obtained with constant $am$.}
\eef{chirenorm}

\bef
\begin{center}
\includegraphics[scale=0.65]{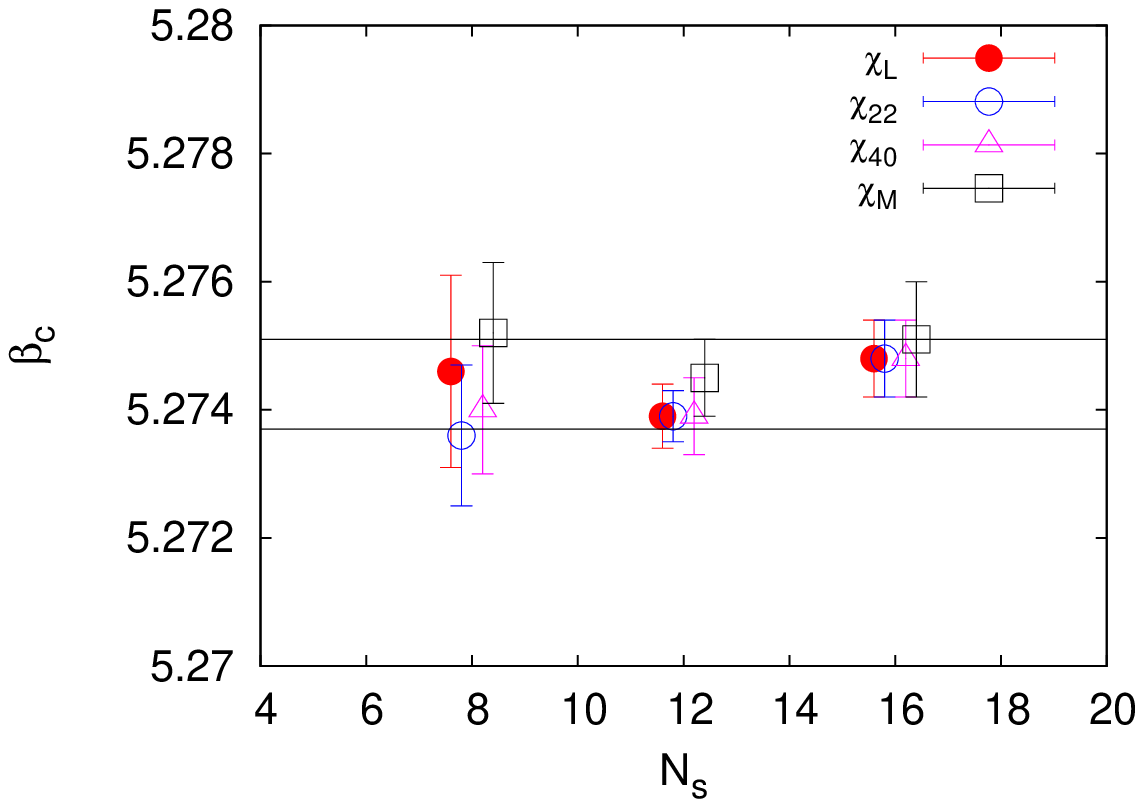}
\includegraphics[scale=0.65]{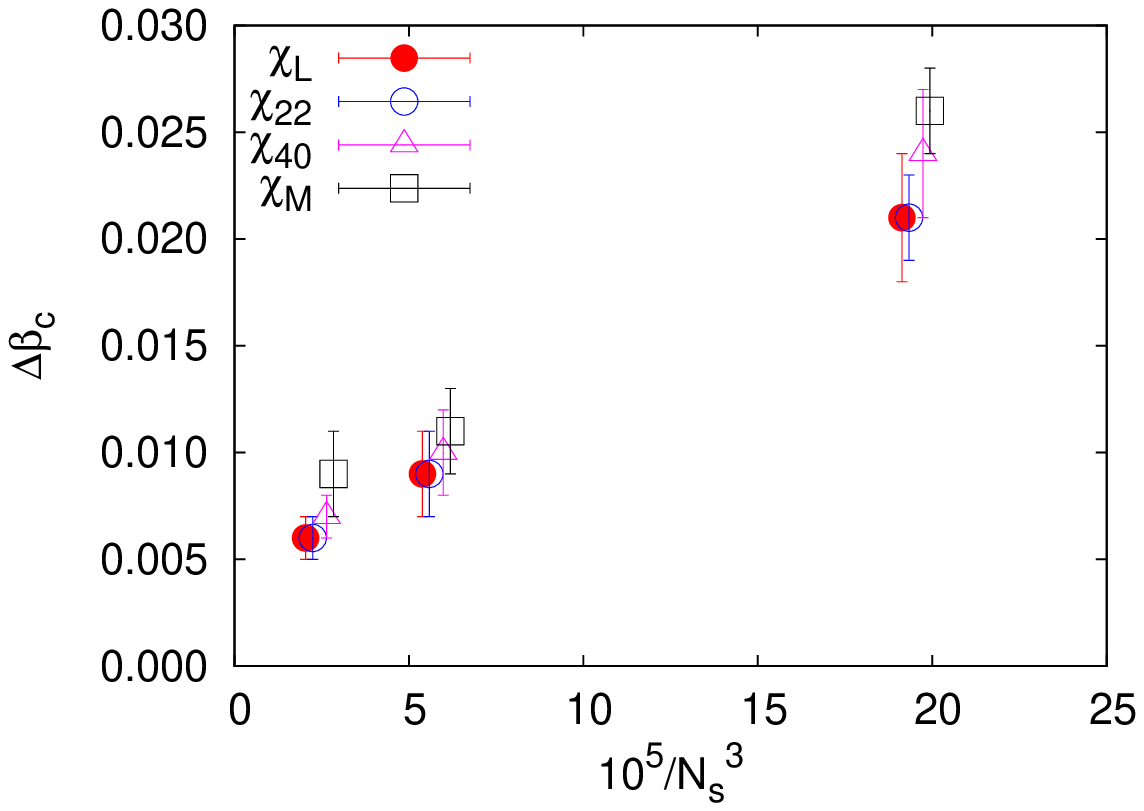}
\end{center}
\vspace{-0.7cm}
\caption{The spatial size dependence of $\beta_c$ (left) and $\Delta\beta_c$
 (right).  The lines enclose 68\% confidence limits on $\beta_c$,
 obtained by fitting a single constant to all the estimates.}
\eef{betavol}

\bet
\begin{center}
\begin{tabular}{|l|c|c|}
\hline
 & $\beta_c$ & $\Delta\beta_c$ \\
\hline
$\chi_\M$ & 5.2747(6) & 0.009(2) \\
$\WLS$ &  5.2743(5) & 0.006(1) \\
$\chi_{22}$ & 5.2741(5)& 0.006(1)\\
$\chi_{40}$& 5.2743(6) & 0.007(1)\\
\hline
\end{tabular}
\end{center}
\caption{$\beta_c$ and $\Delta\beta_c$ as determined from different
 susceptibilities. $\Delta\beta_c$ is much larger than the statistical
 error in $\beta_c$. }
\eet{betas}

The cross over is determined at $N_t=4$ for a bare quark mass
$a_cm=0.015$, where $a_c$ is the lattice spacing at $\beta_c$.
We determined $\beta_c$ by positions of the peaks of different
susceptibilities.  $\Delta\beta_c$ was defined to be the full width at
half maximum (FWHM) of the same susceptibilities.

We measured the Wilson line susceptibility, $\WLS$ \cite{nt6},
the bare chiral susceptibility, $\chi_\M$ \cite{laermanntherm}, the
corresponding renormalized quantity $m_r^2\chi_\M^r/T^4$ \cite{fodor1},
and the fourth order QNS, $\chi_{22}$ and $\chi_{40}$ \cite{nt4}, at
various values of $\beta$ in the crossover region.  For the measurement
of $m_r^2\chi_\M^r/T^4$, we determined the chiral condensate at zero
temperature on $16^4$ lattice at the same values of $\beta$ as the finite
temperature ones.

Prior to the runs listed in \tbn{confnew}, we performed a series of
runs at fixed bare quark mass, $am=0.015$, with $N_s=8$ and 12. We
used these runs to make first estimates of $\beta_c$, and followed up
with the runs along lines of constant $m/T_c$ listed in \tbn{confnew}.
The compatibility of these runs is shown in \fgn{chirenorm}, where
$m_r^2\chi_\M^r/T^4$ is given as a function of $T/T_c$. The figure also
shows that with this cutoff, the deconfining and chiral cross overs in
QCD coincide; $m_r^2\chi_\M^r/T^4$ peaks between 0.98 and $1.02T_c$.

To determine $\beta_c$ accurately, we interpolated data for
susceptibilities using multihistogram reweighting \cite{swnsn} in
the cross over region. From bootstrap resampling of the histograms,
we determined the means and errors in the position of the peak of each
susceptibility and its FWHM, so obtaining $\beta_c$ and $\Delta\beta_c$
\cite{largeNc}. We found $\beta_c$ and $\Delta\beta_c$ for each of the
susceptibilities on the three different lattice volumes. The results
of which are shown in \fgn{betavol}. Since we found very little volume
dependence in $\beta_c$, we made a fit to a constant, independent
of volume.  The values of $\beta_c$ so determined are displayed in
\tbn{betas}. In \fgn{betavol} we also show the volume dependence
of $\Delta\beta_c$.  This decreases with the volume, and gives some
indication of saturating, within errors, close to our largest lattice. So
we take $\Delta\beta_c$ obtained on $N_s=16$, as our best estimate. These
estimates are also listed in \tbn{betas}.  We find that the variation in
$\beta_c$ with different susceptibilities occur well within the width of
the cross over measured from each indicator separately.  In fact, the
four estimates of $\beta_c$ are consistent with each other within 68\%
confidence limits. Combining all four measurements, we quote $\beta_c=
5.2744(7)$ and $\Delta\beta_c\approx0.006$.


\begin{thebibliography}{99}
\bibitem{qcdft}
  J.\ P.\ Blaizot \etal, \plt B 523, 143 (2001);\\
  A.\ Vuorinen, \prd 67, 074032 (2003);\\
  Y.\ Schroder and M.\ Laine, hep-lat/0509104.
\bibitem{kogut}
  C.\ E.\ Detar and J.\ B.\ Kogut, \prl 59, 399 (1987).
\bibitem{fftmass}
  K.\ D.\ Born \etal, \prl 67, 302 (1991).
\bibitem{quasi}
  D.\ Banerjee \etal, \prd 83, 074510 (2011).
\bibitem{cheng}
  M.\ Cheng \etal, {\sl Eur.\ Phys.\ J.\/} C 71, 1564 (2011).
\bibitem{edwin}
  E.\ Laermann and F.\ Pucci, {\sl Eur.\ Phys.\ J.\/} C 72, 2200 (2012).
\bibitem{brandt}
  B.\ B.\ Brandt \etal, arxiv:1210.6972.
\bibitem{dr}
  M.\ Laine and M.\ Veps\"al\"ainen, \jhep 0402, 004 (2004).
\bibitem{htl}
  W.\ M.\ Alberico \etal, \np A\  792, 152 (2007).
\bibitem{scrnp}
  S.\ Datta \etal, arxiv:1212.2927.
\bibitem{APE}
  M.\ Albanese \etal, \plt B\  192, 163 (1987).
\bibitem{HYP}
  A.\ Hasenfratz\ and\ F.\ Knechtli, \prd 64, 034504 (2001).
\bibitem{ST}
  C.\ Morningstar\ and\ M.\ J.\ Peardon, \prd 69, 054501 (2004).
\bibitem{HEX}
  S.\ Capitani, S.\ D\"urr, C.\ Hoelbling, \jhep  0611, 028 (2006).
\bibitem{taste}
  K.\ Orginos, D.\ Toussaint and R.\ L.\ Sugar, \prd 60, 054503 (1999);\\
  E.\ Follana \etal, eprint hep-lat/0406021;\\
  T.\ Bae \etal, \prd 77, 094508 (2008).
\bibitem{nt4}
  R.\ V.\ Gavai and S.\ Gupta, \prd 71, 114014 (2005).
\bibitem{nt6}
  R.\ V.\ Gavai and S.\ Gupta, \prd 78, 114503 (2008).
\bibitem{golub}
  G.\ H.\ Golub and C.\ F.\ van Loan, {\sl Matrix Computations\/},
   Johns Hopkins University Press (1996).
\bibitem{golterman}
  M.\ F.\ L.\ Golterman, \np B 273, 663 (1986).
\bibitem{condit}
  T.\ Kurth \etal, arxiv:1011.1780.
\bibitem{lacaze}
  R.\ V.\ Gavai, S.\ Gupta and R.\ Lacaze, \prd 77, 114506 (2008).
\bibitem{oldlacaze}
  R.\ V.\ Gavai, S.\ Gupta and R.\ Lacaze, \prd 65, 094504 (2002).
\bibitem{ohno}
  H.\ Ohno \etal, arxiv:1211.2591.
\bibitem{tscale}
  S.\ Gupta, \prd 64, 034507 (2001).
\bibitem{laermanntherm}
  E.\ Laermann, \npps 63, 114 (1998).
\bibitem{fodor1}
  Y.\ Aoki \etal, \pl B  643, 46 (2006).
\bibitem{swnsn}
  A.\ M.\ Ferrenberg and R.\ H.\ Swendsen, \prl 63, 1195 (1989).
\bibitem{largeNc}
  S.\ Datta and S.\ Gupta, \prd 80, 114504 (2009).
\end{thebibliography}
\end{document}